\documentclass[twocolumn,showpacs,showkeys,preprintnumbers,nofootinbib]{revtex4}
\usepackage{amsmath,amssymb}

\usepackage{graphicx}
\usepackage{bm}
\usepackage{dcolumn}
\usepackage[mathscr]{eucal}
\usepackage{color}
\usepackage{pst-plot}
\newcommand{\be}{\begin{eqnarray}}
\newcommand{\ee}{\end{eqnarray}}

\renewcommand{\texttt}{{}}

\begin{document}

\title{Spinning Loop Black Holes}
\author{Francesco Caravelli}
\affiliation{Perimeter Institute for Theoretical Physics, 31 Caroline St., Waterloo, ON, N2L 2Y5, Canada}
\affiliation{Department of Physics and Astronomy, University of Waterloo, Waterloo, ON, N2L 3G1, Canada}
\author{Leonardo Modesto}
\affiliation{Perimeter Institute for Theoretical Physics, 31 Caroline St., Waterloo, ON, N2L 2Y5, Canada}
\date{\small\today}

\begin{abstract} \noindent
In this paper we construct four Kerr-like spacetimes starting from the loop black hole Schwarzschild solutions (LBH) and applying the Newman-Janis transformation. 
In previous papers the Schwarzschild LBH was obtained replacing the Ashtekar connection with holonomies on a particular graph in a minisuperspace approximation which describes the black hole interior. 
Starting from this solution, we use a Newman-Janis transformation and we specialize to two different and natural complexifications inspired
from the complexifications of the Schwarzschild and Reissner-Nordstr\"{o}m metrics. 
We show explicitly that the space-times obtained in this way are singularity free and thus there are no naked singularities.
We show that the transformation move, if any, the causality violating regions of the Kerr metric far from $r=0$ .
We study the space-time structure with particular attention to the horizons shape. 
We conclude the paper with a discussion on a regular Reissner-Nordstr\"{o}m black hole derived from
the Schwarzschild LBH and then applying again the Newmann-Janis transformation.
\end{abstract}
\maketitle
\tableofcontents
\section{Introduction}
The quantization of gravity is one of the still open issues in theoretical physics. Several researchers with several different 
approaches are trying to achieve
this target. Loop quantum gravity \cite{LQG} is the most conservative one from this point of view, being based on the Dirac 
quantization and not relying on any exotic idea. However in this context it is difficult to study the semiclassical regime of 
the theory and the classical limit. While several progresses have been made in the context of spin foams, still lot
of work is needed  in order to control the theory \cite{SF}. However, one of the successes of general relativity is its geometric content and it could
be useful to keep a way to deal with the \emph{quantum} using old and well known techniques.


In the context of polymeric quantization this point of view is truly the dominant one. 
In Loop Quantum Black Holes inspired by Loop Quantum Cosmology \cite{Bojowald}, for instance, 
the quantum relies on a strong energy 
condition violating effective stress-energy tensor. 
The same techniques of Loop Quantum Cosmology have been used in the context of Polymeric Black Holes where we have a rich literature 
\cite{Modesto:2008im}, \cite{poly}, \cite{GP}. In this paper in particular we concentrate on the regular
 Schwarzschild metric that has been found in \cite{Modesto:2008im} within the minisuperspace approximation. This metric has several 
interesting properties, first of all the resolution of the singularity, the non expected self-duality property and the stability of the Cauchy horizon.
It would be then interesting at this point to go a step forward and study black holes which possesses an asymptotic 
notion of angular momentum. Within GR such black hole has been found by Roy Kerr in the sixties and today it is well known as the Kerr 
spacetime; its eletrically charged counterpart 
is known as the Kerr-Newman solution. However these solutions are known to be not physical inside to horizon for the presence, as for instance in the Kerr case,
of time machines close to the ring singularity. Being such, we 
can ask if such behavior persists when addressing the resolution of the Kerr ring singularity problem in the context of Loop Black 
Holes. Unfortunately it is known that finding
such type of solution is not an easy task. In the context of loop black holes not even some sort of simplified equations are known. 
Thus we are facing the exceptional task of finding a solution for a spacetime without even 
having the differential equations to 
solve in order to find it. Being such the situation we can try to use a trick to find a solution that would be otherwise impossible 
to obtain. Such trick has been known for 
long time as the Newman-Janis transformation in general relativity. As we will see one of the steps of this transformation, the complexification step, is 
totally arbitrary. However, with a particular choice of the complexification, 
the algorithm gives both the Kerr and the Kerr-Newman solution starting from the Schwarzschild and the Reissner-Nordstr\"{o}m 
respectively. It is then not surprising that when applied to the polymeric Schwarzschild black 
hole the solutions become the ordinary Kerr metric if we put the polimeric parameters and/or $l_P$ 
(the Planck length) to zero. In this paper 
we will use the Newman-Janis algorithm to construct four different rotating spacetimes. 

In this paper we consider two spherically symmetric space-time that we call 
semi-polymeric and full-polymeric.  
In \cite{Modesto:2008im} LBH solutions were obtained replacing the Ashtekar connection with holonomies on a particular graph in a 
minisuperspace approximation which describes the black hole interior. 
We call semi-polymeric the solution which is polymeric 
only in the radial component of the connection 
and full-polymeric the solution obtained replacing all the connection components with holonomies.
The two LBH {\em solutions} are :
\begin{widetext}
\be 
&& {\rm SEMI-POLYMERIC} :  \,\,\,\, \,\,\, ds^2= 
\frac{r^3  \left(r- 2m \right)}{r^4 + a_o^2} dt^2 - \frac{dr^2}{\frac{r^3  \left(r- 2m \right)}{r^4 + a_o^2} }
- \left( r^2 + \frac{a_o^2}{r^2} \right)
d \Omega^{(2)}  \, , \nonumber \\
&& {\rm FULL-POLYMERIC} : \,\,\,\,\,\,
ds^2 = \frac{(r-r_+)(r-r_-)(r+ r_{*})^2}{r^4 +a_o^2} \, dt^2 - \frac{dr^2}{\frac{(r-r_+)(r-r_-) r^4}{(r+ r_{*})^2 (r^4 +a_o^2)}} - \left( r^2 + \frac{a_o^2}{r^2} \right) d \Omega^{(2)} \, ,
\ee
\end{widetext}
where we used signature: $(+, -, -, -)$.
In section II and IV we will give a few more details about the solutions.

We will start from the semi-polymeric metric and choose the two most natural 
complexification 
of the metric compatible with the complexification of the Schwarzschild metric. 
We will discuss the same procedure in the full-polymeric case.

With the risk of being repetitive, the second aim of the paper is the following and is here stressed. 
The Kerr solution is considered as non physical inside the horizon due to the presence of closed time-like curves (CTC) when the metric 
is geodetically extended in the negative radial regions. For this reason Penrose long ago considered the possibility of a \textit{cosmic 
censorship} avoiding the creation 
of naked singularities in general relativity. Being that the CTC's are close to the ring singularity of the Kerr metric, it is 
interesting to study geometries that are 
asymptotically Kerr-like but resolve the ring singularity problem. Recently Smailagic and Spallucci 
 \cite{spallucci}
found the equivalent of the Kerr black hole in the scenario of
noncommutative inspired black holes introduced by Nicolini and Spallucci \cite{NCBH}. Such solution has no ring singularity, no 
superluminal motion and no CTC's and so can be considered as a physical solution. The ring 
singularity is replaced by a rotating classical string which puts in rotation the spacetime. It is then interesting for us to study 
the space-times obtained from the LBH with the NJ algorithm and check if still there are
causality violating regions. 
 

The structure of the paper is the following: in section II we briefly recall the properties of the Schwarzschild loop black hole; in 
section III we review the Newman-Janis transformation; in section IV we apply the
transformation to the semi-polymeric Schwarzschild loop black hole and we study the CTC's for these two metrics; in section V we apply 
the transformation to the full-polymeric Schwarzschild metric. In section VI we introduce the electric charge in the spherically 
symmetric LBHs and we apply again the Newmann-Janis transformation to  
construct the Kerr-Newmann space-time; conclusions follow. In the paper we use natural units $c=G=\hbar=1$.

\section{The regular Schwarzschild-metric}
\label{static}

Let us first summarize the regular black hole metric \cite{Modesto:2008im} that will be
the starting point in the following.
The solution 
is obtained from the canonical quantization of the Einstein 
equations written in terms of the Ashtekar variables, that is in terms of an $\mathrm{SU}(2)$ 3-dimensional connection $A$ and 
a triad $E$. The result is that the basis states of LQG are closed graphs the edges of which are labelled by 
irreducible $\mathrm{SU}(2)$ representations and the vertices by $\mathrm{SU}(2)$ intertwiners. Physically, the edges represent quanta 
of area with area $\gamma l_{\rm P}^2 \sqrt{j(j+1)}$, where $j$ is the representation label on the edge (a half-integer), $l_{\rm P}$ is 
the Planck length,  and $\gamma$ is a parameter of order 1 called the Immerzi parameter. Vertices of the graph represent quanta of 
3-volume. The important observation to make here is that area is quantized and the smallest quanta of area possible has area 
$\sqrt{3}/2 \gamma l_{\rm P}^2$. 

The regular black hole metric that we will be using is derived from a simplified model of LQG \cite{Modesto:2008im}. To obtain this 
simplified model we make the following assumptions. First of all, the number of variables is reduced by assuming spherical symmetry. 
Then, instead of all possible closed graphs, a regular lattice with edge lengths $\delta_b$ and $\delta_c$ is used. The solution is then 
obtained dynamically inside the homogeneous region (inside the horizon where space is homogeneous but not static). Analytically 
continuing the solution outside the horizon one finds that one can reduce the two free parameters by imposing that the minimum area 
presents in the solution corresponds to the minimum area of LQG. The one remaining unknown constant $\delta_b$ is a parameter of the 
model determining the strength of deviations from the classical theory, and would have to be constrained by experiment. With the 
plausible expectation that the quantum graviational corrections become relevant only when the curvature is in the Planckian regime, 
corresponding to $\delta_b < 1$, outside the horizon the solution is the Schwarzschild solution up to negligible Planck-scale 
corrections which allows us to believe the legitimacy of the analytical extension outside the horizon. The analytical extension is 
supported by a rigorous analysis explained in detail 
in \cite{Modesto:2008im}.

This quantum gravitationally corrected
Schwarzschild metric can be expressed in the form
\begin{eqnarray}
ds^2 = G(r) dt^2 - \frac{dr^2}{F(r)} - H(r) d\Omega^2,
\label{g}
\end{eqnarray}
with $d \Omega = d \theta^2 + \sin^2 \theta d \phi^2$ and
\begin{eqnarray}
&& G(r) = \frac{(r-r_+)(r-r_-)(r+ r_{*})^2}{r^4 +a_o^2}~ , \nonumber \\
&& F(r) = \frac{(r-r_+)(r-r_-) r^4}{(r+ r_{*})^2 (r^4 +a_o^2)} ~, \nonumber \\
&& H(r) = r^2 + \frac{a_o^2}{r^2}~ .
\label{statgmunu}
\end{eqnarray}
Here, $r_+ = 2m$ and $r_-= 2 m P^2$ are the two horizons, and $r_* = \sqrt{r_+ r_-} = 2mP$. $P$ is the
polimeric function $P = (\sqrt{1+\epsilon^2} -1)/(\sqrt{1+\epsilon^2} +1)$, with
$\epsilon \ll 1$ the product of the Immirzi parameter ($\gamma$) and the polimeric parameter ($\delta_b$). With this, it is 
also $P \ll 1$, such that $r_-$ and $r_*$ are very close to $r=0$. The area $a_o$ is equal to $A_{\rm min}/8 \pi$, $A_{\rm min}$ being 
the minimum area gap of LQG.

Note that in the above metric, $r$ is only asymptotically the usual radial
coordinate since $g_{\theta \theta}$ is not just $r^2$. This choice of
coordinates however has the advantage of easily revealing the properties
of this metric as we will see. But first, most importantly, in the limit
$r \to \infty$ the deviations from the Schwarzschild-solution are of
order $M \epsilon^2/r$, where $M$ is the usual ADM-mass:
\be
G(r) &\to& 1-\frac{2 M}{r} (1 - \epsilon^2)~, \nonumber  \\
F(r) &\to& 1-\frac{2 M}{r}~ , \nonumber \\
H(r) &\to& r^2 .
\ee
The ADM mass is the mass inferred by an observer at flat asymptotic infinity; it is determined solely 
by the metric at asymptotic infinity.  The parameter $m$ in the solution is related to the mass $M$ by $M = m (1+P)^2$.

If one now makes the coordinate transformation $R = a_o/r$ with the rescaling 
$\tilde t= t \, r_*^{2}/a_o$, and
simultaneously substitutes $R_\pm = a_o/r_\mp$, $R_* = a_o/r_*$ one finds that the metric in
the new coordinates has the same form as in the old coordinates and thus exhibits a
very compelling type of self-duality with dual radius $r=\sqrt{a_o}$. Looking at the angular part
of the metric, one sees that this dual radius corresponds to a minimal possible 
surface element. It is then also clear that in the limit $r\to 0$, corresponding
to $R\to \infty$, the solution
does not have a singularity, but instead has another asymptotically flat Schwarzschild region.

The metric in Eq. (\ref{statgmunu}) is a solution of a quantum gravitationally corrected set of
equations which, in the absence of quantum corrections $\epsilon, a_o \to 0$, reproduces Einstein's field equations.

\section{The Newman-Janis algorithm}
In this section we review the Newman-Janis transformation for a generic spherically symmetric spacetime \cite{DSNJ}. Roughly speaking,
the algorithm start from a non-rotating spacetime and at the end of the steps the spacetime is rotating. We stress that one of the steps 
is arbitrary but can be constrained by the classical limit. The starting point a spherically symmetric spacetime.
In its most general form, the metrics are of the following form, 
\be
&& ds^2=e^{2\Phi(r)}dt^2-e^{2\lambda(r)}dr^2 - H(r)  d\Omega^2 \nonumber \\
&& \hspace{0.6cm} := G(r) dt^2-\frac{dr^2}{F(r)}-H(r) d\Omega^2,
\label{GSS}
\ee
which define the function $\Phi(r)$ and $\lambda(r)$ used in literature with $G(r)$ and $F(r)$. 
The first step of the transformation is to change coordinates. This step requires the advanced null coordinates 
$\left\{u,r,\vartheta, \phi\right\}$, where 
\be
u=t-r^\ast
\ee
and $dr^\ast= dr/\sqrt{G F}$. The line element above then becomes,
\be
ds^2 = G(r) du^2 + 2 \sqrt{\frac{G(r)}{F(r)}} \,  du dr - H(r) \, d\Omega^2 ,
\label{GFHmetric}
\nonumber
\ee
while the non zero components of the inverse metric are 
\be
&& g^{u \phi} =  e^{- \Phi(r) - \lambda(r)} \,\,\, , \,\,  g^{\phi \phi} = - [ H(r) \sin^2 \theta ]^{-1} , \nonumber \\
&& g^{\theta \theta} = - H(r)^{-1} \,\,\, , \,\, g^{r r} = - e^{- 2 \lambda(r)}.
\ee

The second step of the algorithm is to find the null tetrads for the inverse matrix as follows,
\be
&& g^{\mu \nu } = l^{\mu} n^{\nu} +  l^{\nu} n^{\mu} -  m^{\mu} {\bar m}^{\nu} -  m^{\nu} {\bar m}^{\mu} \nonumber\\
&& l^{\mu } = \delta_r^{\mu} \nonumber \\
&& n^{\mu} = \sqrt{\frac{F}{G}} \delta^{\mu}_u - \frac{1}{2} F \delta^{\mu}_r, \nonumber \\
&& m^{\mu} = \frac{1}{\sqrt{ 2 H}} \left( \delta_{\theta}^{\mu} + \frac{ i}{\sin \theta} \delta_{\phi}^{\mu} \right). 
\label{tetra}
\ee
where the vectors satisfy the relations
$l_{\mu} l^{\mu} = m_{\mu} m^{\mu} = n_{\mu} n^{\mu} = l_{\mu} m^{\mu} =n_{\mu} m^{\mu} =0$
and $l_{\mu} n^{\mu} = - m_{\mu} {\bar m}^{\mu} =1$ (${\bar x}$ is the complex conjugate of the general quantity $x$). 
The main step of the procedure is the combination of two operations. The first complex transformation in the $r-u$ plane as follows:
\begin{eqnarray}
&& r \rightarrow r^{\prime} = r + i \, a \cos \theta,  \nonumber \\
&&  u \rightarrow u^{\prime} = u  - i \, a \cos \theta.
\label{NJ}
\end{eqnarray}
together with a complexification of the functions $F$,$G$ and $H$ of the metric, under which the null tetrads become
\be
&& \hspace{-0.6cm} l^{\mu } = \delta_1^{\mu} \nonumber \\
&& \hspace{-0.6cm} n^{\mu} = \sqrt{\frac{\tilde F(r')}{\tilde G(r')}} \, \delta^{\mu}_u - \frac{1}{2} \tilde F(r')\,  \delta^{\mu}_r,  \\
&& \hspace{-0.6cm} m^{\mu} = \frac{1}{\sqrt{ 2 \tilde H(r')}}  
\left(i a \sin \theta (\delta^{\mu}_u - \delta^{\mu}_r) + \delta_{\theta}^{\mu} + \frac{ i}{\sin \theta} \delta_{\phi}^{\mu} \right). \nonumber 
\label{tetra2}
\ee
where $\tilde F$, $\tilde G$ and $\tilde H$ are real functions on the complex domain. This step of the procedure is in principle 
completely arbitrary. In fact in the original paper Newman and
Janis could not give a true explanation of the procedure if not that it works for the Kerr metric with a particular choice of the 
complexifications. The situation improved with Drake and Szekeres \cite{DSNJ}, 
in which they prove that the only Petrov D space-times generated by the Newman-Janis algorithm with a vanishing Ricci scalar is the 
Kerr-Newman spacetime. Having this fact in mind, the non zero 
components of the inverse metric (\ref{GFHmetric}) become,
using the tetrads (\ref{tetra2}) after the transformation,
%
%
\be
&& g^{u u } = -\frac{a^2 \sin^2(\theta )}{ \tilde{H} (r ,\theta )} \,\,\, , \,\,
g^{u \phi} = -\frac{a}{ \tilde{H} (r ,\theta )} , \nonumber  \\
&& g^{\phi \phi} = -\frac{1}{  \tilde{H} (r, \theta ) \sin^2 \theta } \,\,\, , \,\, 
g^{\theta \theta} = -\frac{1}{ \tilde{H} (r,\theta )},  \nonumber \\
&& g^{rr} = -\frac{a^2 \sin ^2 \theta }{  \tilde{H} (r, \theta)} - e^{- 2\lambda(r, \theta)} \,\,\, , \,\, 
g^{r \phi} = \frac{a}{\tilde{H} (r, \theta)} ,
\nonumber  \\
&& g^{u r } = \frac{a^2 \sin ^2(\theta )}{ \tilde{H} (r, \theta)} + e^{- \tilde \Phi (r, \theta) - \tilde \lambda(r, \theta)}.
\label{KCinvm}
\ee

Let us now apply the procedure to 
the classical Schwarzschild example 
In this case the metric has $G =F$ or $\lambda = - \Phi$ and $H=r^2$. The metric reads, in usual Eddington-Finkelstein coordinates:
\be
&& ds^2 = G(r) du^2 + 2dudr -r^2 d\Omega^2 \nonumber \\ 
&& = \left(1 - \frac{2 m}{r} \right) du^2 + 2dudr -r^2 d\Omega^2
\nonumber
\ee
where $G(r)=1-\frac{2m}{r}$ and we see that $H(r)=r^2$. If we apply the Newman-Janis algorithm as prescribed above, we have to choose a complexification of the 
$r^2$ and of the $1/r$ term. In general this prescription is not
unique. However since we know what the Kerr solution is, we know that if we take the following complexification:
\be
&& r^2 \rightarrow r' \bar{r}'  , \nonumber \\
&& \frac{1}{r} \rightarrow \frac{1}{2} \left(\frac{1}{r'}+\frac{1}{\bar{r}'} \right).
\label{schwarz}
\ee
This trick works well. This is the same as complexifying in the following way the functions $G(r)$ and $H(r)$
\begin{eqnarray}
&& \tilde G(r')=1-m \left(\frac{1}{r'}+\frac{1}{\bar r'} \right) = 1-\frac{2m r}{r^2+a^2 \cos^2 \theta} , \nonumber \\
&& \tilde H(r')=r' \bar r' = r^2 + a^2 \cos^2 \theta.
\label{schwarzschild}
\end{eqnarray}
in the classical Schwarzschild metric. 
The find metric is the Kerr metric in Kerr-Schild coordinates. However let us stress that in practice nothing could have told us, 
without using the Einstein equations, 
that a particular complexification
is favored respect to the other one. This situation is even worse for Reissner-Nordstr\"{o}m. In fact in this last case the function 
$G(r)$ is of the form:
\begin{equation}
 G(r)=1-\frac{2m}{r}-\frac{Q^2}{r^2},
\label{rng}
\end{equation}
where Q is the electric charge of the black hole. In this case the two most natural complexification of the last term in (\ref{rng}) is:
\begin{equation}
\frac{1}{r^2}\rightarrow \frac{1}{r' \bar r'}.
\end{equation}
After this introduction we can apply the procedure to the LBHs.
\section{Semi-polymeric spinning LBH}
The line element for the semi-polymeric black hole can be obtained from (\ref{statgmunu}) setting $r_{*}=r_{-}=0$ and depends only on 
the mass
and the minimum area $a_o \propto l_P^2$ ($l_P = \sqrt{G_N \hbar}$ is the Planck length)
(but can be obtained also from a simpler Hamiltonian constraint \cite{Modesto:2008im})
\be 
&& ds^2= 
\frac{r^3  \left(r- 2m \right)}{r^4 + a_o^2} dt^2 - \frac{dr^2}{\frac{r^3  \left(r- 2m \right)}{r^4 + a_o^2} }
- H(r) 
d \Omega^{(2)} , \nonumber \\
&& H(r) =  r^2 + \frac{a_o^2}{r^2} . 
\ee
This metric is regular everywhere and reproduces the Schwarzschild metric in the limit $a_o \rightarrow 0$.
We can work with a more general form of the metric leaving the physical radius of the two sphere
implicit
\be
ds^2= 
\frac{r^2  \left(1- \frac{2m}{r} \right)}{H(r)} dt^2 - \frac{dr^2}{\frac{r^2  \left(1- \frac{2m}{r} \right)}{H(r)}}
- H(r) d \Omega^{(2)}. 
\ee
and in Eddington-Finkelstein coordinates because $G = F$ (or $\Phi = - \lambda$) reduces to 
\be 
&& ds^2=
 \frac{r^2  \left(1- \frac{2m}{r} \right)}{H(r)} du^2+2dudr-H(r) d\Omega^2 \nonumber \\
&& \hspace{0.7cm} =  \frac{r^2  \left(1- \frac{2m}{r} \right)}{r^4 + a_o^2} du^2+2dudr-H(r) d\Omega^2.
\label{semigen} 
\ee
Now we complexify the functions $G$ and $H$ appearing in the metric. It is easy to understand that, in $G$, the term of the form 
$(1- 2m/r)$ must be
complexified as (\ref{schwarz}) for compatibility with the Schwarzschild metric in the limit $a_o,a\rightarrow0$. On the same footage 
the $r^2$ term in $H(r)$, for
compatibility with Kerr in the $a_o\rightarrow 0$ limit. This means that in $G$ the $r^2$ must be complexified as $r^2\rightarrow r' \bar r'$ for compatibility with the Kerr metric
in the limit $a_o\rightarrow0$. Thus we are left only with the complexification of the $a_o$ term in $H$, which represent the 
\emph{quantum} correction of the metric. The two most natural complexification 
of the term proportional to $a_{o}$ in $H(r)$ are:
\begin{equation}
{\rm Type \,\, I} \, : \,\,\,\,\,\,  \frac{a_o^2}{r^2} \rightarrow \frac{a_o^2}{(r'+\bar r')^2/4}
\label{type1}
\end{equation}
or, as the Reissner-Nordstr\"{o}m case suggests:
\begin{equation}
\hspace{-1cm} {\rm Type \,\, II} \, : \,\,\,\,\,\,   \frac{a_o^2}{r^2} \rightarrow \frac{a_o^2}{r' \bar r'}
\label{type2}
\end{equation}
In the following we refer to complexification (\ref{type1}) as Type I and (\ref{type2}) as Type II.
\subsection{Type I complexification} \label{TypeI}
\begin{figure}
 \begin{center}
  \includegraphics[height=5.5cm]{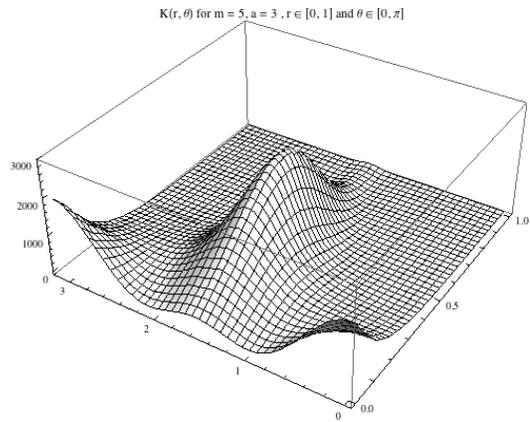}
  \end{center}
  \caption{\label{K2a} Plot of the Kretschmann invariant for $(m,a) = (5,3)$ and $r \geqslant 0$ 
  in Planck units. Type I complexification.}
  \end{figure}

\begin{figure}
 \begin{center}
      \includegraphics[height=5.5cm]{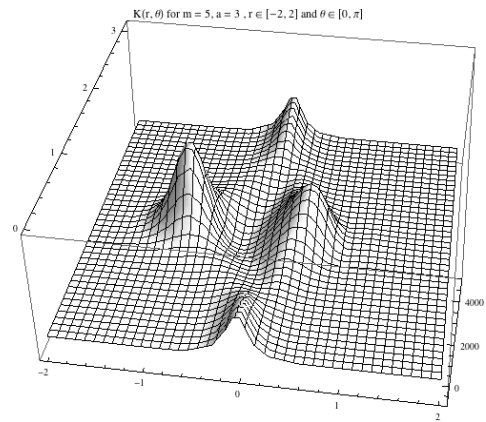}
  \end{center}
  \caption{\label{K2b} Plot of the Kretschmann invariant for $(m,a) = (5,3)$ and positive and negative values
  of $r$ 
  in Planck units. Type I complexification.}
  \end{figure}

As explained earlier we proceed to complexify the components of the metric as
\be
&&  G(r) = \frac{r^2 \left(1 - \frac{2 m}{r}  \right)}{H(r)}     
\mapsto 
\tilde{G}(r^{\prime}) := G(r, \theta) , \nonumber \\
&&
G(r, \theta) = r' \bar{r}' \,
 \left[1 - 2 m \, \frac{1}{2}\left(\frac{1}{r'} + \frac{1}{\bar{r}'} \right)  \right]  \, \frac{1}{\tilde{H}(r')} 
 \nonumber \\
 &&  \hspace{1.15cm} = 
  \left( \rho^2 - 2 m r \right)   \,\frac{1}{\tilde{H}(r')}  =
  \frac{\left( \rho^2 - 2 m r \right)}{\Sigma},  \nonumber \\
&& H(r) = r^2 + \frac{a_o^2}{r^2}  \mapsto \tilde{H}(r') := \Sigma(r, \theta)  , \nonumber \\
&& 
\Sigma(r, \theta) 
= r' \bar{r}' + \frac{a_o^2}{(r'+\bar{r}')^2/4} = 
\rho^2 + \frac{a_o^2}{r^2},  \nonumber \\
&& \hspace{0cm} \rho^2(r, \theta) := r^2 + a^2 \cos^2 \theta . 
\label{complexGH2}
\ee
in the original Kerr coordinates the metric reads 
\be
&& ds^2_{(K)} = G(r, \theta)  du^2 - \Sigma(r , \theta) d \theta^2 - 2 a \sin^2 \theta dr d \phi  \nonumber \\
&& + \left[ a^2 (G(r,\theta )-2) \sin ^2 \theta - \Sigma (r,\theta ) \right] \sin ^2(\theta ) d \phi^2 \nonumber \\
&& + 2 a (1-G(r,\theta )) \sin ^2 \theta  d \phi \, d u .
\label{core2}
\ee
In Kerr coordinates the metric is regular everywhere contrary to the classical one. There is no singularity on the event horizons and 
is simpler to show the regularity of this space-time.
 \begin{figure}
 \begin{center}
  \includegraphics[height=5.5cm]{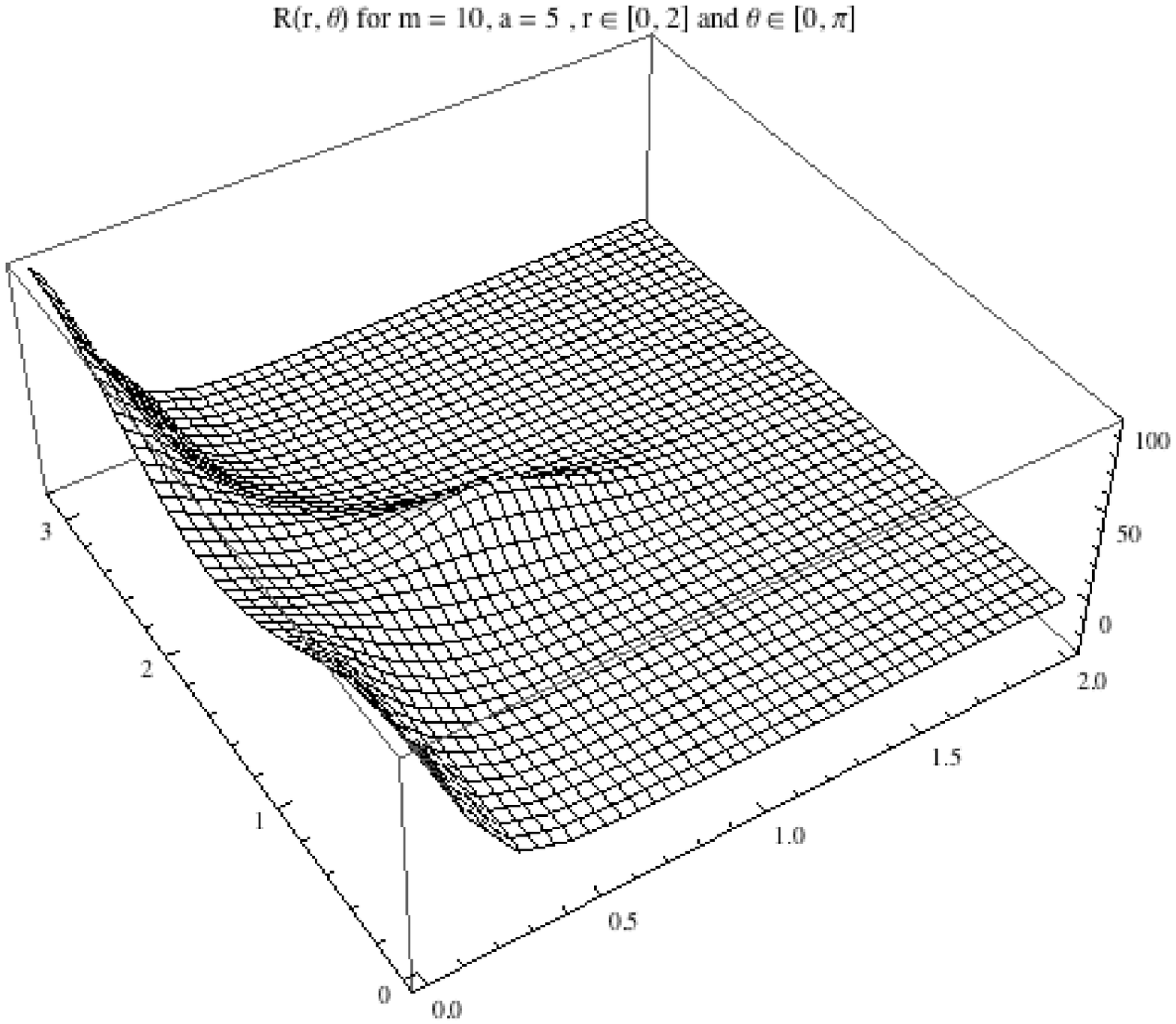}
    \includegraphics[height=6.0cm]{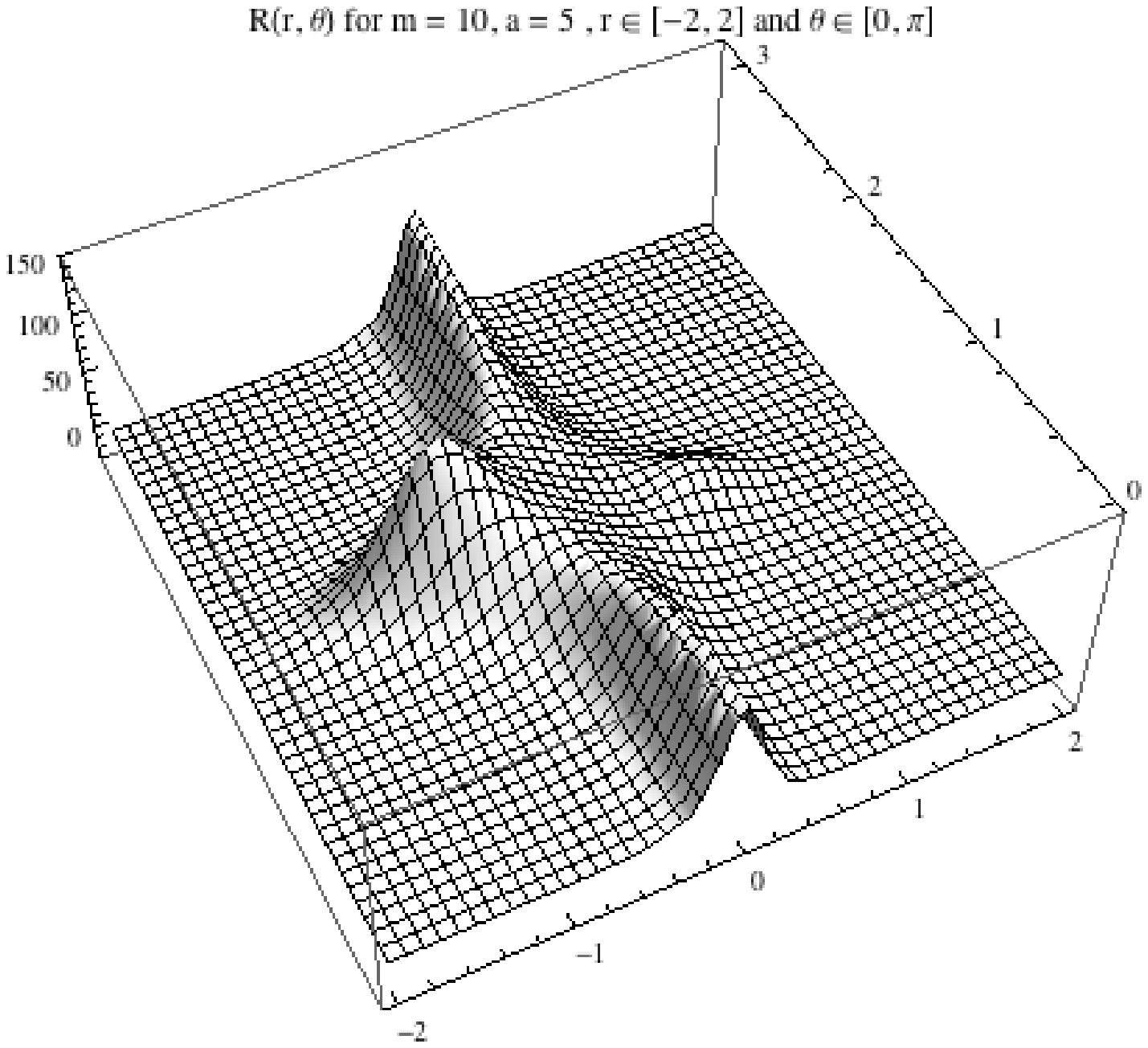}
  \end{center}
  \caption{\label{RicciPlot} Plot of the Ricci scalar for $m=10$, $a=5$ when the radial coordinate assume 
  only positive values and for $r \in [-2 , 2]$ in the second plot. Type I complexification.}
  \end{figure}
  By a coordinate transformation the metric can be written in Boyer-Lindquist coordinates 
\be
&& ds^2_{B-L} = G(r , \theta)  dt^2  
- \frac{\Sigma (r,\theta )  \, dr^2 }{a^2 \sin ^2 \theta +G(r,\theta ) \Sigma (r,\theta ) } \nonumber \\
&&       +2  (1 - G(r,\theta )) \sin ^2 \theta  \, d t d \phi        -\Sigma (r,\theta ) d \theta^2
 \nonumber \\
&& 
- \sin ^2 \theta  \left[ a^2 (2 - G(r,\theta )) \sin ^2 \theta  +  \Sigma (r, \theta )\right] d \phi^2
\ee
In order to simplify the notation we introduce the following quantities 
\be
&& \Delta(r) = G (r, \theta) \Sigma(r, \theta) + a^2 \sin^2 \theta = r^2 - 2 m r +a^2, \nonumber \\
&& G(r, \theta) = \frac{\Delta(r) - a^2 \sin^2 \theta}{\Sigma(r , \theta)} ,
\label{delta2}
\ee
inside the metric and we write down the line element explicitly in Boyer-Lindquist (B-L) coordinates 
defined by the coordinate transformation $du = dt + g(r) d r$, $d \phi = d \phi' + h(r) dr$ 
(we will omit the dependents on $\theta$ and $r$ in the function $\Delta$, $\Sigma$)
where 
\be
&& g(r) = - \frac{ e^{\lambda} (   \Sigma +a^2 e^{\lambda + \Phi} )}{e^{\Phi} ( \Sigma + a^2 \sin^2 \theta e^{2 \lambda} )} , \nonumber \\
&& h(r) = - \frac{ a e^{2 \lambda}}{\Sigma + a^2 \sin^2 \theta e^{2 \lambda}}, 
\label{BLtransf}
\ee
are valid for general functions $\Phi$ and $\lambda$. The B-L metric reads 
\be
&&  \hspace{-0.7cm}  ds^2 =  \frac{\Delta - a^2 \sin^2 \theta}{\Sigma} dt^2 - \frac{ \Sigma}{\Delta }  \, dr^2    \nonumber \\
&& \hspace{-0.7cm} 
 +  2 a \sin^2 \theta \left(1 - \frac{\Delta - a^2 \sin^2 \theta}{\Sigma} \right) dt \, d \phi  
 - \Sigma \, d \theta^2 
 \nonumber \\
&&  \hspace{-0.7cm} 
 -  \, \sin ^2 \theta  \left[ \Sigma + a^2 \sin^2 \theta \left(2 - \frac{\Delta - a^2 \sin^2\theta}{\Sigma}\right)   \right]    d \phi^2.
%
%
  \label{SPK2}
\ee
The first quantity we study is the Ricci scalar tensor which is classically zero in the empty space 
but not now, because of the quantum geometry effects,
\be
&& \hspace{-1cm} R(r,\theta) = \frac{8 a_o^2}{\left(a^2 r^2 \cos (2 \theta
   )+a^2 r^2+2 a_o^2+2 r^4\right)^3} \times \nonumber \\
&&  \hspace{-1.0cm}
 \Big[  3 a^4 r^2+a^2 \cos (2 \theta ) \left(3 r^2 \left(a^2-2 m r\right)+a_o^2\right) \nonumber \\
&&  \hspace{-1cm}
+3 a^2
   \left(a_o^2-2 m r^3+6 r^4\right)-4 r \left(a_o^2+3 r^4\right) (2 m-r)  \Big],
\label{Rscalar}
\ee
which is zero for $a_o \rightarrow 0$. On the equatorial plane ($\theta = \pi/2$)
and $r = 0$, $R (0, \pi/2) = 2 a^2/a_o^2$.
The plots in Fig.\ref{RicciPlot} show the Ricci scalar 
is non singular and peaked in $\theta \approx \pi/2$, $r \approx \sqrt{a_o}$.
To complete the singularity resolution analysis we should analyze the regularity 
properties of the Kretschmann invariant tensor
$K := R_{\mu \nu \rho \sigma}  \, R^{\mu \nu \rho \sigma}$. 
This quantity is given in the appendix 
but the regularity properties are shown with the tool of
the two dimensional plots. The plots 
are given in Fig.\ref{K2a},\ref{K2b} 
where it is evident the metric is regular everywhere and 
the value of the curvature in $r = 0$, $\theta = \pi/2$ is $K(0, \pi/2) = 4 a^4/a_o^4$.

The component $g_{tt}$ of the metric changes sign on the surfaces defined by
\be
g_{tt} = 0 \,\,\, \rightarrow \,\,\, \Delta(r) - a^2 \sin^2 \theta = 0,
\ee
or more explicitly in 
\be
r = m \pm \sqrt{ m^2 - a^2 \cos^2 \theta},
\ee
that define the ergosphere of the classical Kerr space-time.
  \begin{figure}
 \begin{center}
  \hspace{-0.5cm}
     \includegraphics[height=10cm]{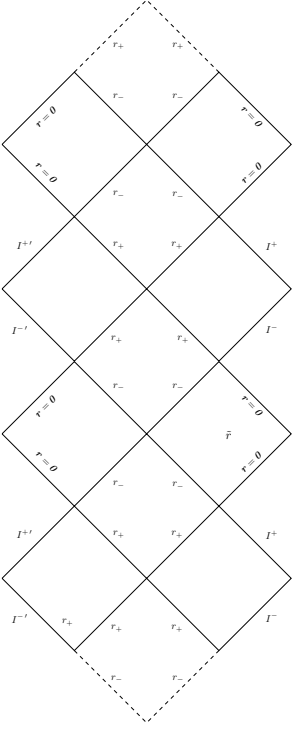}
  \end{center}
  \caption{\label{penSPI} Penrose diagram for $\theta = 0$ and $r \geqslant 0$.}
  \end{figure}

   The horizons are those surfaces 
   where the light cones lies and then also the light can not escape.
   In other words they are defined by 
   \be
   \left(\frac{d r}{d t} \right)^2 = 0 \,\,\, \forall \,\,\, d \theta \, , \,\,  d \phi.
   \label{horizons}
   \ee
   In our case this happens only where $\Delta(r)  = 0$ or 
   \be
   r = r_{\pm} := m \pm \sqrt{ m^2 - a^2} \,\,\, {\rm for} \,\,\,  a < m .
   \ee
   The event horizon is a null surface and a Killing surface as we are going to show.
   The surface $\Sigma(t, r,\theta, \phi) = {\rm const}.$
is a null surface if the normal $n_i = \partial \mathcal{S}/ \partial x^i$ is a null vector 
or satisfies the condition $n_i n^i =0$. 
The last identity says that the vector $n^i$ is on the surface $\mathcal{S}(t, r, \theta, \phi)$ itself,
in fact $d  \mathcal{S} = d x^i \partial \mathcal{S}/\partial x^i$ and $d x^i \| n^i$.  The norm of
the vector $n_i$ is 
\begin{eqnarray}
n_i n^i = g^{ij} \frac{\partial \mathcal{S} }{\partial x^i} \frac{\partial \mathcal{S} }{\partial x^i} = 0.
\label{NullS}
\end{eqnarray}
In our case (\ref{NullS}) reduces to 
\begin{eqnarray}
g^{rr} \frac{\partial \mathcal{S} }{\partial r} \frac{\partial \mathcal{S}  }{\partial r} +
g^{\theta \theta} \frac{\partial \mathcal{S}  }{\partial \theta} \frac{\partial \mathcal{S}}{\partial \theta} =0.
\end{eqnarray}
and this equation is satisfied where $g^{rr}(r) = 0$ if the surface 
is independent from $\theta$, $\mathcal{S}(r, \theta) = \mathcal{S}(r)$. The points where 
$g^{rr} =0$ are $r_{+}$ and $r_{-}$ and $r = 0$ but only $r_-$ and $r_+$ are horizons.  
The metric admits two killing vector $t^{\mu} = \partial_t$ and $\phi^{\mu} = \partial_{\phi}$
but also any linear combination is a Killing vector.
In particular 
\be
\xi^{\mu} = t^{\mu} + \Omega_H \phi^{\mu},
\label{KV}
\ee
  is a Killing vector and for 
  \be
  \Omega_H = \frac{a r_{+}^2}{r_{+}^2 (a^2 + r_{+}^2) + a_o^2}
  \ee 
   it is null on the event horizon ($r_+$), $\xi^{\mu} \xi_{\mu}|_{r_+} = 0$; 
   this concludes the proof that the event horizon is a Killing horizon.
   We can calculate also the surface gravity on $r_{+}$ and $r_{-}$. It is defined in
   terms of the Killing vector (\ref{KV}) by
   \be
   \kappa^2 = - \frac{1}{2} \nabla^{\mu} \xi^{\nu} \nabla_{\mu} \xi_{\nu}
   \ee
   and the result is
   \be
   && \kappa_{+} = \frac{r_{+}^2 (r_+ - r_{-})}{2 \left(a^2 r_{+}^2+a_o^2+r_{+}^4\right)}, \nonumber \\
   && \kappa_{-} = \frac{r_{-}^2 (r_{+}-r_{-})}{4 \left(a^2 r_{-}^2+ a_o^2+r_{-}^4\right)}.
   \ee
  \begin{figure}
 \begin{center}
  \hspace{0.0cm}
     \includegraphics[height=8cm]{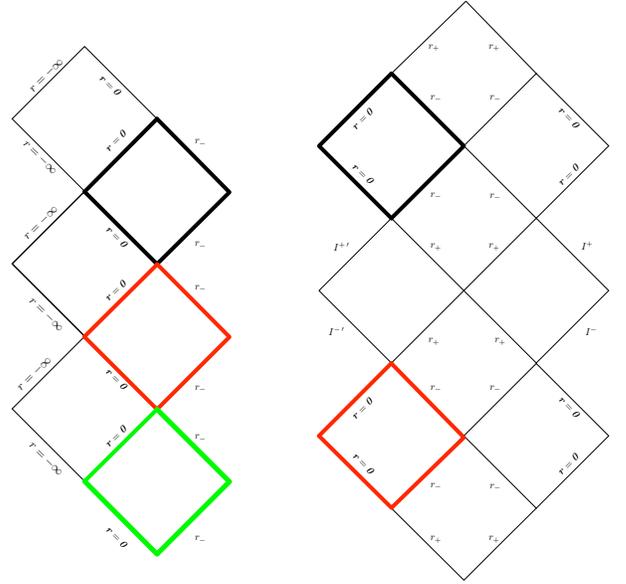}
  \end{center}
  \caption{\label{penSPIM} Maximal extension of the Type I space-time for $a< m$ and 
  $\theta = 0$. The surface $r =0$
  is a null surface but it is not an horizon beside $g_{tt}$ and $g_{rr}$ do not change sign. Block 
  of the same color has to be identify to have the maximal extension of the space-time to negative value 
  of $r$. }
  \end{figure}
To conclude the event horizon area is 
\be
A_H = 4 \pi \frac{ (r_+^4 + a^2 r_+^2 + a_o^2)}{r_+^2}.
\label{AH}
\ee
   The metric for $\theta$ and $\phi$ constant reduces to 
\be
ds^2 =  \frac{(\Delta - a^2 \sin^2 \theta_o)}{\Sigma} dt^2- \frac{ \Sigma}{\Delta }  \, dr^2 
\label{S2D2}
\ee
 and the tortoise coordinate for $\theta = 0$ is 
\be
&& \hspace{-0.25cm}
r^* = r -\frac{a_o^2}{r \, r_- r_+}  +\frac{a_o^2 \left( r_- +  r_+\right)}{r_-^2
   r_+^2} \log |r|  \nonumber \\
   && \hspace{0.8cm}
  + \frac{\left(a^2 r_-^2+a_o^2+r_-^4\right) }{r_-^2
   (r_- - r_+)} \log |r - r_-|  \nonumber \\
   && \hspace{0.8cm} 
   + \frac{\left(a^2 r_+^2  + a_o^2 + r_+^4\right)}{r_+^2
   (r_+ - r_-)}  \log |r-r_+| . 
   \label{tortoise}
   \ee
  We can introduce first the coordinates $u = t - r^*$ and $v = t + r^*$ and then 
  $U^{\pm} = \mp \exp( \mp \kappa_{\pm} u)/\kappa_{\pm}$,
   $V^{\pm} = \pm \exp( \pm \kappa_{\pm} v)/\kappa_{\pm}$ 
   for $r> r_-$ and $r< r_-$ respectively. Looking to 
   $U^- V^- = - \exp(- 2 \kappa_- r^*)/\kappa_-^2$ we see that 
   $U^- V^- \rightarrow 0$ for $r \rightarrow r_-$ and 
   $U^- V^- \rightarrow - \infty$ for $r =0$. 
   The Penrose diagram
   in Fig.\ref{penSPI} is a Penrose diagram for $r>0$. Despite the position 
   of the point $r =0$ in the diagram it is not an event horizon
   as can be seen solving (\ref{horizons}). A maximal extension to negative
   values of $r$ is obtained following the analysis in \cite{Walker}.
  The result is given in Fig.\ref{penSPIM} .
    
\begin{figure}
 \begin{center}
  \hspace{-0.5cm}
     \includegraphics[height=9cm]{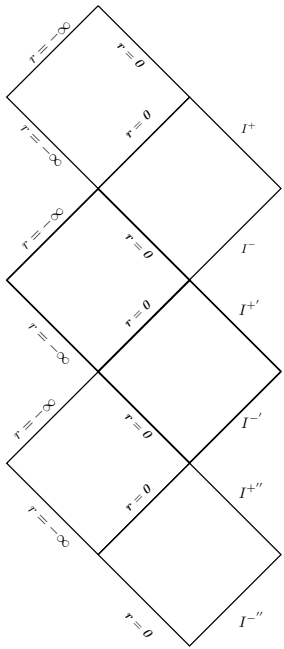}
  \end{center}
  \caption{\label{PenamagmI} Maximal extension of the Type I space-time for $a> m$ and 
  $\theta = 0$. The surface $r =0$
  is a null surface but it is not an horizon beside $g_{tt}$ and $g_{rr}$ do not change sign.}
  \end{figure}

\begin{figure}
 \begin{center}
  \hspace{-0.5cm}
     \includegraphics[height=9cm]{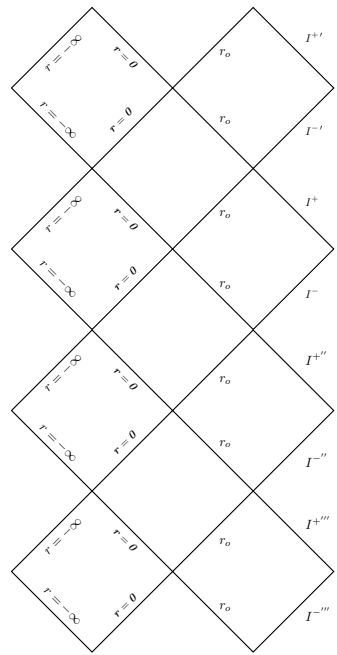}
  \end{center}
  \caption{\label{PenaegualmI} Maximal extension of the Type I space-time for $a = m$ and 
  $\theta = 0$ ($r_o = m$).}
  \end{figure}

   
  As we showed studying the Ricci scalar and in particular the Kretschmann invariant
  the space-time is regular everywhere. If we plot the Kretschmann invariant 
  for the case $a > m$ we obtain plots similar to those in 
   Fig.\ref{K2a},\ref{K2b},\ref{penSPI}. In other words we do not have naked singularities.
   The tortoise coordinate for $a>m$ and $\theta = 0$ is 
   \be
&&   r^* = r  -\frac{a_o^2}{a^2 r}   + \frac{2 a_o^2 m \log | r | }{a^4} \nonumber \\
   &&
   +\frac{\left(2 a^4 m^2-a^2 a_o^2+2 a_o^2 m^2\right)
   \arctan \left(\frac{r-m}{\sqrt{a^2-m^2}}\right)}{a^4 \sqrt{a^2-m^2}} \nonumber \\
  && +\frac{m \left(a^4-a_o^2\right)}{a^4} \log
   \left(a^2-2 m r+r^2\right)
   \label{tortoiseabigm}
   \ee
   In this case, to understand the causal structure of the space-time,
    we can also introduce coordinates $(u,v)$ and then a single couple 
   of new coordinates $(U, V)$ because there is just one coordinate singularity in $r = 0$. 
   The result is a block of space-time which extend from $+ \infty$ to $r=0$.
   Following again \cite{Walker} the maximal extension of the space-time is given in
   Fig.\ref{PenamagmI}.
For the extremal case $m = a$ and $\theta = 0$
the tortoise coordinate is
\be
&& r^* = r -\frac{a_o^2}{r r_o^2}  -\frac{a^2 r_o^2+a_o^2+r_o^4}{r_o^2 (r-r_o)}+\frac{2 a_o^2 }{r_o^3}  \log | r | \nonumber \\
   && \hspace{1cm}
   +  \frac{2 \left(r_o^4-a_o^2\right)}{r_o^3}  \log  | r-r_o |
\ee
 and the Penrose diagram is in Fig.\ref{PenaegualmI}.
  
Following the analysis of the classical Kerr metric \cite{Walker} we have focused our attention 
to the axis of symmetry because it is easier to study. Nevertheless it seems probable 
again as in the classical case that the basic topological properties of the four dimensional 
space-time are essentially the same.

\subsection{Type II complexification} \label{TypeII}
In this section we consider a different complexification starting from the 
same spherically symmetric metric (\ref{semigen}) 
in Eddington-Finkelstein coordinates 
\be 
ds^2=\frac{r^2  \left(1- \frac{2m}{r} \right)}{H(r)} du^2+2dudr-H(r) d\Omega^2.
\ee

\begin{figure}
 \begin{center}
  \includegraphics[height=6cm]{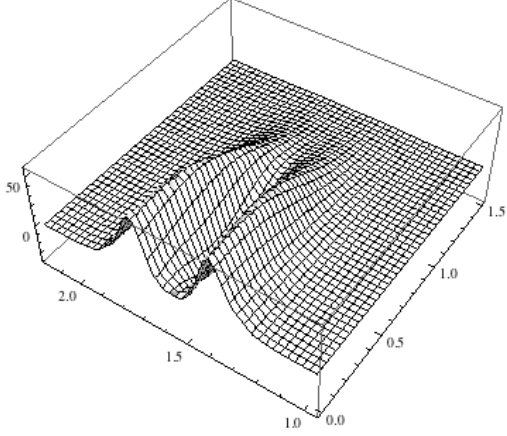}
  \end{center}
  \caption{\label{RicciII} Plot of the Ricci scalar for $m=10$, $a=5$ in Planck units ($r>0$).}
  \end{figure}

\begin{figure}
 \begin{center}
   \includegraphics[height=6.4cm]{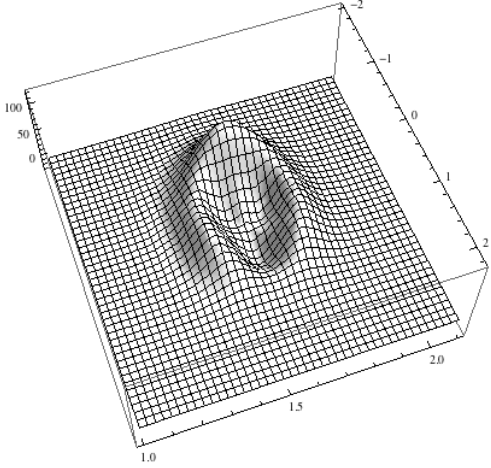}
  \end{center}
  \caption{\label{RicciIIneg} Plot of the Ricci scalar for $m=10$, $a=5$ in Planck units and $r\in ] - \infty, + \infty[$. 
The behavior at the origin is : 
  $\lim_{\theta \rightarrow \pi/2} \lim_{r \rightarrow 0} R(r, \theta) = 
  \lim_{r \rightarrow 0} \lim_{\theta \rightarrow \pi/2} R(r, \theta) =
  2 a^2/a_o^2$.
  }
  \end{figure}

\begin{figure}
 \begin{center}
 \includegraphics[height=5.8cm]{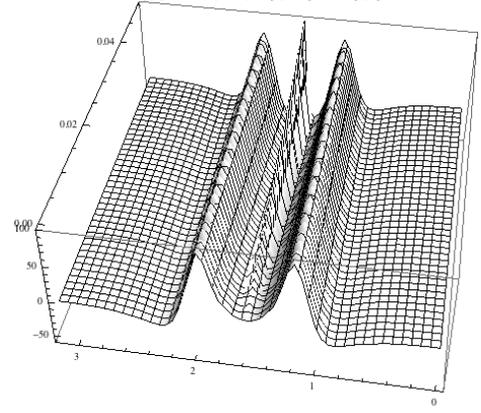}  
  \end{center}
  \caption{\label{KIIA} Type II metric:  Plot of the Kretschmann scalar for $m=3$, $a=2$ in Planck units near $r = 0$.}
  \end{figure}

\begin{figure}
 \begin{center}
    \includegraphics[height=4.8cm]{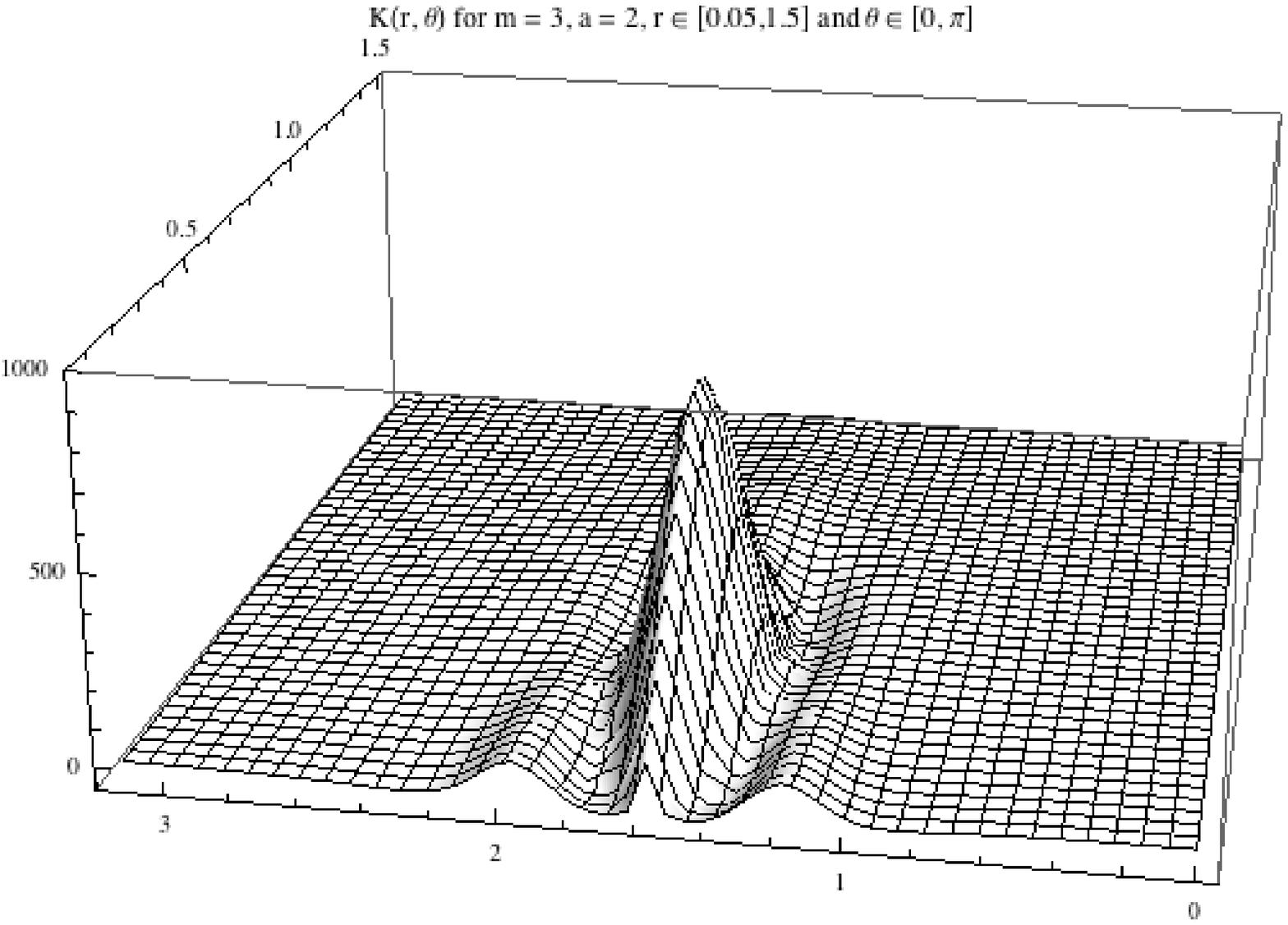}  
  \end{center}
  \caption{\label{KIIB} Type II metric:  Plot of the Kretschmann scalar for $m=3$, $a=2$ in Planck units
  for $r>0$.}
  \end{figure}

\begin{figure}
 \begin{center}
     \includegraphics[height=6.8cm]{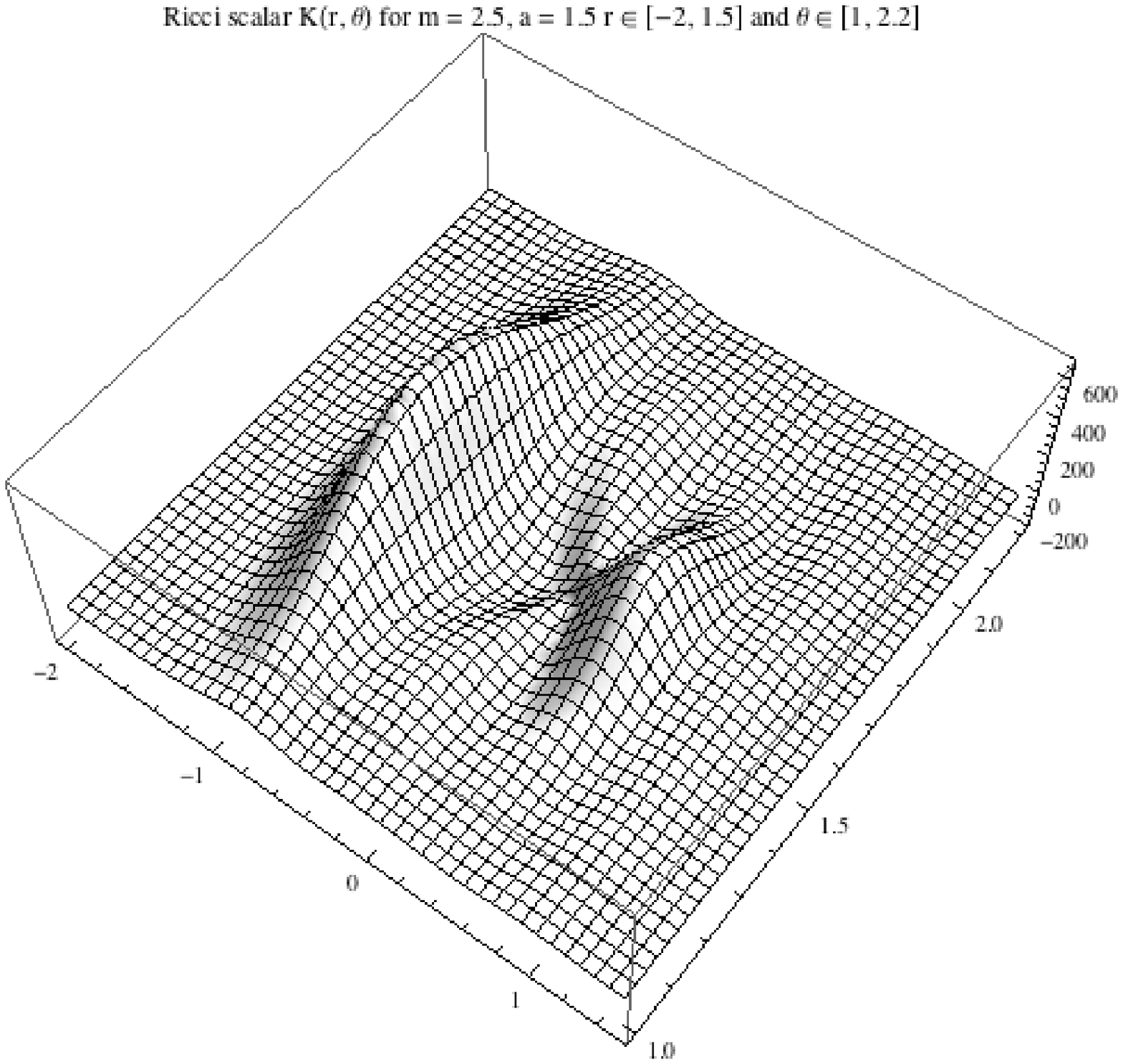}
  \end{center}
  \caption{\label{KIIC} Type II metric:  Plot of the Kretschmann scalar for $m=3$, $a=2$ in Planck units. This plot involves also 
negative values of $r$ and the behavior at the origin is: 
  $\lim_{\theta \rightarrow \pi/2} \lim_{r \rightarrow 0} K(r, \theta) = 
  \lim_{r \rightarrow 0} \lim_{\theta \rightarrow \pi/2} K(r, \theta) = 0$.
  }
  \end{figure}

As explained the complexification 
(\ref{complexGH2}) changes 
 only in the factor $H(r)$,
\be
&& H(r) \mapsto \Sigma(r, \theta) = r' \bar{r}' + \frac{a_o^2}{r' {\bar r}'} = 
\rho^2 + \frac{a_o^2}{\rho^2},  \nonumber \\
&& \hspace{0cm} \rho^2(r, \theta) := r^2 + a^2 \cos^2 \theta.
\label{complexGH3}
\ee
This small modification is sufficient to make the metric harder to study. 
In Kerr coordinates the metric is regular everywhere contrary to the classical one
and is given in (\ref{core2}). There is no singularity on the event horizons and is more simple to show the regularity of the improved 
space-time metric.
Again, the first quantity we study is the Ricci scalar tensor. 
The plots in Fig.\ref{RicciII},\ref{RicciIIneg} show the Ricci scalar 
is non singular. To complete the
 singularity resolution analysis we should analyze the Kretschmann invariant tensor
$K := R_{\mu \nu \rho \sigma}  \, R^{\mu \nu \rho \sigma}$. 
For this metric, more then for the type I, $K(r, \theta)$ 
 is very involved and will show its regularity properties with the tool of
three dimensional plots. The three dimensional plots are in Fig.\ref{KIIA},\ref{KIIB},\ref{KIIC}.
Passing to Boyer-Lindquist coordinates the metric assumes the same form of 
(\ref{SPK2}) but with a different function $\Sigma(r, \theta)$ defined in (\ref{complexGH2}).
Because $\Delta(r)$ is also the same the event horizon and the ergosphere surfaces are the same.
The tortoise coordinate for $\theta = 0$ and $a < m$ is 
\be
&& \hspace{0cm} 
r^* = r + \frac{a_o^2 (r_- + r_+) }{2 \left(a^2+r_-^2\right)
   \left(a^2+r_+^2\right)}
  \log \left(a^2+r^2\right) \nonumber \\
  && \hspace{1cm}
   -\frac{a_o^2  \left(a^2-r_- r_+\right)}{a
   \left(a^2+r_-^2\right) \left(a^2+r_+^2\right)} 
  \arctan \left(r/a \right)
   \nonumber  \\
   && \hspace{1cm} 
   -\frac{\left(a^4+2 a^2 r_-^2+a_o^2+r_-^4\right)}{\left(a^2+r_-^2\right) (r_+ - r_-)} \log |r - r_-| \nonumber \\
   && \hspace{1cm}
   + \frac{\left(a^4+2 a^2
   r_+^2+a_o^2+r_+^4\right)}{\left(a^2+r_+^2\right) (r_+ - r_-)}  \log | r-r_+|.
   \ee
Given the tortoise coordinate and using the same analysis of the previous section for the type I metric
we can obtain the Penrose diagrams for the type II metric. In the $\theta = 0$ case 
the diagrams are exactly the same of the classical Kerr metric for $a<m$, $a=m$ and $a>m$.
There is no naked singularity and for 
$\theta = \pi/2$ the diagram for the case $a>m$
looks like the Minkowski space-time diagram, as in the classical case, but with an extension 
to negative values of $r$. 

\subsection{Closed time-like curves}
The metrics obtained in this section, as shown, have the nice properties of keeping the singularity free property of the original metric. 
Thus we can now see {\em if} the CTCs disappear in such space-times, since the ring
singularity is not present anymore in both cases. In the Kerr case such CTCs are in the negatively $r$ geodetically extended space-time 
sector. 

In order to study the CTC's problem we study the norm of the Killing vector along the $\phi$ direction. 
This vector has norm $\phi^{\mu} \phi_{\mu} = g_{\phi \phi}$. 
We calculate such norm for the classical metric and for the Type I and Type II metrics.
We consider the norm near the point $r = 0$, $\theta = \pi/2$. 
Let $r/a = \delta$ (small and negative) and consider $\theta = \pi/2 + \delta$.
Then classically we find 
\be
\phi_{\mu}\phi^{\mu} = g_{\phi \phi}= -\frac{a m}{\delta } - a^2 + {\mathcal O} \left(\delta \right),
\ee
which is positive for sufficiently small and negative $\delta$.
For the Type I and Type II loop black holes instead we find 
\be
&& \phi_{\mu}\phi^{\mu}  = - \frac{ a_o^2 }{a^2 \delta ^2} + {\rm const.} + {\mathcal O} (\delta),
\nonumber \\
&&   \phi_{\mu}\phi^{\mu}  = - \frac{ a_o^2 }{2 a^2 \delta ^2} + {\rm const.} + {\mathcal O} (\delta)
\ee
that are always negative for small values of $\delta$.
We conclude there are no CTC in the region around $r \approx 0$ and $\theta \approx \pi/2$
contrary to the classical Kerr space-time.

\begin{figure}
 \begin{center}
   \includegraphics[height=6.5cm]{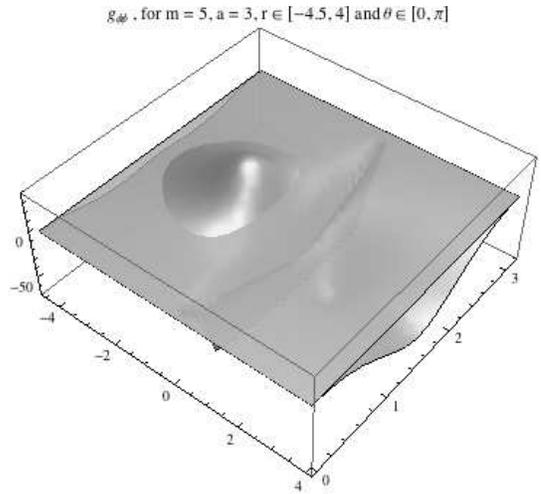}
 \end{center}
 \caption{In this plot the region where $g_{\phi \phi}$ is positive contains CTC curves
 in the analytical extension to $r<0$ but not for $r\rightarrow 0$. This plot refers to the Type I but 
 for the Type II is almost equal.}
\label{ctcs}
  \end{figure}

However for negative values of $r$ and arbitrary values of $\theta$ the norm of the Killing
vector can change sign as showed in Fig.\ref{ctcs} and we can still have CTC curves.
The lump region in the plot is a {\em time machine region} \cite{TM}.

%
%
%

\section{The full-polymeric spinning LBH} \label{FullPolyKerr}
In this section we apply the Newmann-Janis \cite{NJ} transformation to the loop black hole metric in
its full polimeric form, (\ref{statgmunu}). This metric is quite complicated than the 
semi-polymeric one, thus we restrict our analysis to more features independent from the complexification 
of the function $H(r)$. Moreover we restrict ourself to a particular complexification of the functions, even if 
suggested by the complexification of the Reissner-Nordstr\"{o}m metric. The full polimeric line element can be rewritten as follows,
\be
&& \hspace{-0.3cm} ds^2 = g_{tt} dt^2 + g_{rr} dr^2 + g_{\Omega} d \Omega^{(2)}, \nonumber \\
&& \hspace{-0.3cm} g_{tt} := e^{2 \Phi(r)} = \left(1 - \frac{r_+ + r_-}{r}  + \frac{r_+ r_-}{r^2} \right) \left( 1 + \frac{r_*}{r} \right)^2    \hspace{-0.2cm}
 \frac{r^2}{H(r)}, \nonumber \\
&& \hspace{-0.3cm}  g_{rr} := - e^{2 \lambda(r)} = 
- \frac{ \left( 1+\frac{r_*}{r} \right)^2    }{ \left( 1 - \frac{ r_+ + r_-}{r} + \frac{r_+ r_-}{r^2}  \right)  } 
\frac{H(r)}{r^2} ,  \nonumber \\
&& \hspace{-0.3cm} g_{\Omega} := - H(r) = -\left( r^2 +\frac{a_o^2}{r^2} \right), 
\label{components}
\ee
where $r_+ = 2m$ and $r_- = 2 m P^2$ and $r_* = 2 m P$ as already defined in the first section of the paper.
The terms of the form $1/r$, $1/r^2$ and $r^2$ are naturally complexified as the Reissner-Nordstr\"{o}m terms, leaving us the choice, 
as for the semi polimeric case, to 
complexify the function $H(r)$,   
\be
&& \frac{1}{r} \mapsto 2 \left(\frac{1}{r'} + \frac{1}{\bar{r}'} \right) , \nonumber \\
&& \frac{1}{r^2} \mapsto \frac{1}{r' \bar{r}'}, \nonumber \\
&& r^2 \mapsto r' \bar{r}'.
\ee
%
%
%
The line element in Kerr coordinates is 
\be
&& ds^2 = e^{2 \Phi} d u^2 + 2 e^{\lambda + \Phi} du dr   -2 a \sin^2 \theta  e^{\lambda + \Phi} dr d \phi 
 \nonumber \\
&&
- \Sigma d \theta^2 
- \sin^2 \theta \left( \Sigma + a^2 \sin^2 \theta \left( 2 e^{\lambda + \Phi} - e^{2 \Phi} \right) \right)  d \phi^2
\nonumber \\
&& + 2 a \sin^2 \theta \left(  e^{\lambda + \Phi} - e^{2 \Phi} \right) du d \phi , 
\label{FullKerrC}
\ee
where $\Sigma$ comes from the complexification of $H(r)$ and 
\be
&& \hspace{-0.4cm} 
e^{2 \Phi(r, \theta)} = \frac{e^{\lambda(r, \theta) + \Phi(r, \theta)} \,\, \left( \rho^2(r, \theta) - (r_+  +  r_-) r + r_-  r_+   \right)}{   \Sigma(r, \theta)}, \nonumber \\ 
&&  \hspace{-0.4cm} 
e^{\lambda(r, \theta) + \Phi(r, \theta)} = \left( 1 + \frac{r \, r_*}{\rho^2(r, \theta)} \right)^2 
\label{kerrkerr}
\ee
and $\rho^2 := r^2 + a^2 \cos^2 \theta$ is the same function introduced in the semi-polymeric case.

We show now the regularity of the solution considering the Type I complexification of 
$H(r) \rightarrow \Sigma(r, \theta)$.
We can rewrite the metric (\ref{kerrkerr}) in a conformal shape where the conformal factor is
$\exp(\Phi+\lambda) := \exp (2 \sigma)$. The metric reads 
\be
g_{\mu \nu} = e^{\Phi + \lambda} \, \bar{g}_{\mu \nu} := e^{2 \sigma} \, \bar{g}_{\mu \nu}.
\label{confmetric}
\ee
where $\bar{g}_{\mu \nu}$ is regular $\forall \, r \geqslant 0$ 
(this is very simple to see for the Type I complexification because the components 
of the metric $\bar{g}_{\mu \nu}$ never diverge for $\theta = \pi/2$)
and presents the usual bounce 
of the two-sphere in $r = 0$.  
Now we consider the Ricci scalar which can be written in the following way
\be
R=  e^{- 2 \sigma} \, \left(\bar{R} + 6 (\bar{\nabla} \sigma) \right).
\label{RicciFull}
\ee
where $\bar{R}, \bar{\nabla}$ are defined by $\bar{g}_{\mu \nu}$.
When we replace the components of the metric in (\ref{RicciFull}) for $\theta = \pi/2$ we find 
the following leading term 
\be
R \approx e^{- 2 \sigma} \, 6 (\bar{\nabla} \sigma) \approx \frac{6 a^2}{a_o^2}.
\label{RicciFull2}
\ee
which shows the Ricci invariant does not diverges on the equatorial plane for $r=0$.
The behavior of the Ricci scalar (\ref{RicciFull}) 
is a strong argument in favor the regularity of the metric $g_{\mu \nu}$ for $\theta \approx \pi/2$ and 
$r\geqslant 0$. 
Another argument pro regularity of the space-time comes from the radial geodesic 
analysis (for $r \approx 0$) 
in the Boyer-Lindquist coordinates we are going to introduce. We will return on this point
having introduced such coordinates.

The components of the loop improved Kerr metric can be written in Boyer-Lindquist 
coordinates applying the transformation (\ref{BLtransf}). The result is 
\be
&& g_{tt} = \frac{ \left( \Delta(r) + a^2 \cos^2 \theta \right) (\rho^2 + r r_* )^2}{\rho^4 \, \Sigma},
\nonumber \\
&& g_{rr} = - \frac{ \Sigma \, (\rho^2 + r r_*)^2}{\rho^4  \left( \Delta(r) + a^2 \cos^2 \theta \right)
+ a^2 \sin^2 \theta ( \rho^2 + r r_*)^2} , \nonumber \\
&& g_{t \phi} = \frac{a \sin^2 \theta \left( \rho^2 + r r_* \right)^2 \left[ \Sigma - 
( \Delta(r) + a^2 \cos^2 \theta  ) \right]}{\Sigma \, \rho^4} , \nonumber \\
&& g_{\theta \theta} = - \Sigma, \nonumber \\
&& g_{\phi \phi}= -  \sin^2 \theta \Big[ \Sigma \nonumber \\
&& \hspace{-0.2cm} + a^2 \sin^2 \theta \frac{ ( \rho^2 + r r_* )^2 
\left( 2 \Sigma - (\Delta(r) + a^2 \cos^2 \theta  ) \right) }{\Sigma \, \rho^4} \Big],
\label{FPKerr}
\ee
where $\Sigma(r, \theta)$ is the complexification of $H(r)$ and we introduced the notation
\be
\Delta(r) = r^2 - (r_+ + r_-) r + r_+ r_-.
\ee
The ergosphere is quite similar to the classical one and is defined by the surface 
\be
g_{tt} = 0 \,\,\,\, \rightarrow \,\,\,\, \Delta(r) + a^2 \cos^2 \theta =0
\ee
or more explicitly 
\be
&& r = \frac{ r_+ + r_- \pm \sqrt{(r_+  - r_-)^2-4 a^2 \cos ^2 \theta } }{2} \nonumber \\
&&  = m(1 + P^2) \pm \sqrt{ m^2 (1 - P^2)^2 - a^2 \cos^2 \theta}.
\label{ergo1}
\ee
The event horizon is defined by (\ref{horizons}) and such relation for the full polymeric metric reads 
\be
\rho^4  \left( \Delta(r) + a^2 \cos^2 \theta \right)
+ a^2 \sin^2 \theta ( \rho^2 + r r_*)^2 =0.
\label{fullpolyhorizons}
\ee
The definition of black hole horizon we are using here is the following: it is a surface within all lightlike paths 
and hence all paths in the forward light cones of particles within the horizon, are warped so as to fall farther into the hole. Equation (\ref{fullpolyhorizons}) defines a null surface as it is easy to see.

\begin{figure}
 \begin{center}
  \hspace{-0.5cm}
  \includegraphics[height=6cm]{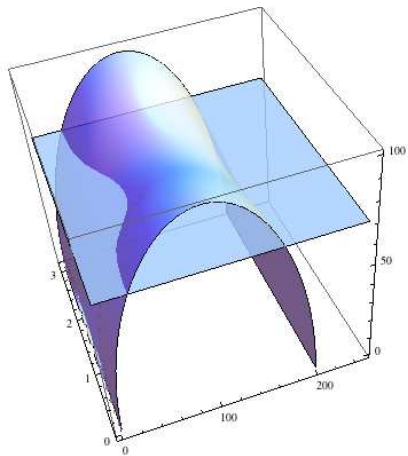}
  \end{center}
  \caption{\label{Horizons1} Contour-plot for the geometric surface where 
  equation (\ref{fullpolyhorizons}) is satisfied. The intersecting plane is $a={\rm const.}$.
  This plot refers to the case of two event horizon one inside the other for $m=100$ in Planck units and $P=0.1$.}
  \end{figure}

\begin{figure}
 \begin{center}
  \hspace{-0.5cm}
  \includegraphics[height=6cm]{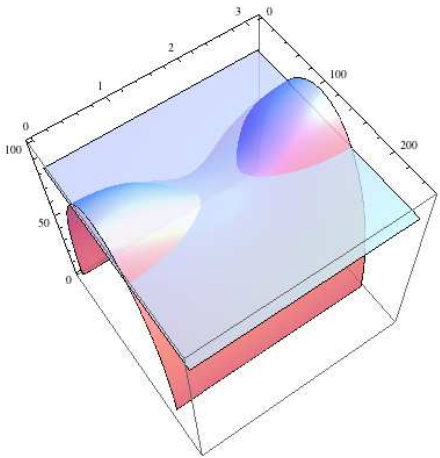} 
  \end{center}
  \caption{\label{Horizons0} Contour-plot for the geometric surface where 
  equation (\ref{fullpolyhorizons}) is satisfied. The intersecting plane is $a={\rm const.}$.
  This plot shows the new phenomenon of two distinct horizons for $m=100$ in Planck units and $P=0.1$.}
  \end{figure}

Contour-plots of the six order equation (\ref{fullpolyhorizons}) are given in Fig.\ref{Horizons1},\ref{Horizons0}.
On the $x,y$ axes are $r$ and $\theta$ on the $z$ axes is the angular momentum $a$.
The horizontal plane for constant $a$ shows explicitly that for small values of $a$ we have two 
quasi-spherical event horizons (Fig.\ref{Horizons0}) one inside the other but for a sufficient large $a$ we have two horizons separated from each other (Fig.\ref{Horizons1}). For the second configuration 
the following 
picture follows: from an observer outside the hole the black hole splits along the 
symmetry axes in two distinct black holes both with ellipsoidal horizon.
In the presence of two horizons, one inside the other, or zero horizons 
the Penrose diagrams (for $\theta = 0$) are the same of Type I or Type II 
respectively 
depending on 
the complexification of $H(r)$. 
In the case of two topologically distinct event horizon the Penrose diagram representation 
is not well define. Indeed, in this case we have two black hole each with a single event horizon
but both inside the same ergosphere. 
\begin{figure}
 \begin{center}
  \hspace{-0.5cm}
  \includegraphics[height=6cm]{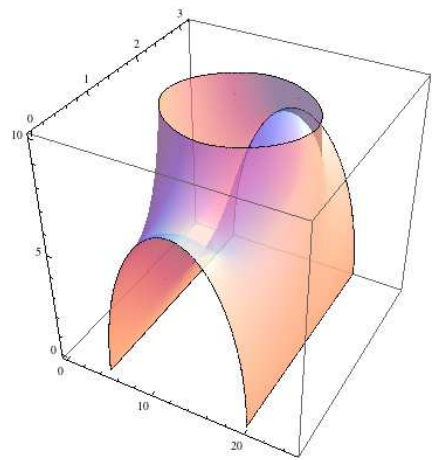} 
  \end{center}
  \caption{\label{fullhorizon} Contour-plot for the geometric surface where the event horizon 
  equation (\ref{fullpolyhorizons}) is satisfied together with the ergosphere surface. The event horizon surface
  is always inside the ergosphere surface and the two surface meet at the poles $\theta=0, \pi$.
  In this plot $m=10$ in Planck units and $P=0.5$ but it is true for any value of the parameters.}
  \end{figure}
%

Another elegant way to verify the non existence of singularity in $r=0$ for $\theta = \pi/2$ 
is to study the geodesic in the equatorial plane.
We have a singularity problem only if the proper time to arrive in $r=0$ is finite.
For the metric (\ref{FPKerr}) on the plane $\theta = \pi/2$ 
orbits are parametrized by the conserved energy per unit of mass, $E$, and the angular momentum 
per unit of mass along the symmetry axis  
the radial geodesic 
of a massive particle can be obtained from the norm of the $4-$velocity and the 
conserved quantities associated 
with the symmetries of the metric 
\be
&& E =  t^{\mu} \, U^{\nu} g_{\mu \nu} \,\, , \,\,\,\, \ell = - \phi^{\mu} \, U^{\nu} g_{\mu \nu} \, , \nonumber \\
&& U^{\mu} \, U_{\mu} = 1,  
\label{GEODIF}
\ee
where we recall the Killing vectors $t^{\mu} =(1,0,0,0)$ and $\phi^{\mu} = (0,0,0,1)$.
The second equation on (\ref{GEODIF}) for general value of $E$, $\ell$ and small value of $r$
reads
\be
\frac{1}{2} \left( \frac{dr}{d \tau} \right)^2 \approx -\frac{a^2 r^2}{2 a_o^2} - \frac{ (a E \ell + r_- r_+ )}{2 a_o^2 r_*^2} \, r^4,
\ee
since the leading term is negative for any value of $E$ and $\ell$ we conclude there is a barrier 
which prevent to arrive in $r=0$ both for positive and negative values or $r$.
In the classical case a test particle can arrive in $r=0$ only from 
$r<0$ because $\dot{r}^2 \propto - 1/r^5$ ($\dot{r}$ is the derivative respect to the proper time $\tau$). Any test particle 
will move around the {\em ring} ($r=0$, $\theta = \pi/2$) without ever reach it. 
Moreover, for $\theta = 0$ any particle arrive in $r=0$ in finite time and then is natural
an analytical extension to negative values of the coordinate $r$.

This is a strong argument in favor of the regularity of the space-time in $r=0$ but not exclude the possibility 
to have singularities for negative values of $r$. We can indeed to reach the region $r<0$ 
starting from $\theta \neq \pi/2$.
This result is independent from the Type I or Type II complexification because 
for $\theta = \pi/2$ the two different complexification of $\Sigma$ coincides.
On the other side 
for $\theta \neq \pi/2$ there is no singularity because the presents of the angular momentum 
$a$ in $\Sigma$ and in $\rho^2$.  

\subsection{Violation of causality}
We showed in the previous section that any observer on the equatorial plane never  
reaches the origin $r = 0$.
This result is important in relation to the CTCs because those curves
exist only for negative values of $r$ which in this metric are not part of the maximal extension
for $\theta = \pi/2$. We can conclude that there are no CTCs for an observer arrives in $r=0$ 
traveling on the plane 
$\theta = \pi/2$ because the maximal extension does not involve $r<0$.
What about the region $r \approx 0$ and $\theta \approx \pi/2$?
We consider the Killing vector along the $\phi$ direction. This 
vector has norm $\phi^{\mu} \phi_{\mu} = g_{\phi \phi}$. 
We compare such norm in the classical case with the new regular metric. 
Let $r/a = \delta$ (small) and consider $\theta = \pi/2 + \delta$.
Then classically 
\be
\phi_{\mu}\phi^{\mu} = g_{\phi \phi}= -\frac{a m}{\delta } - a^2 + {\mathcal O} \left(\delta \right),
\ee
which is positive for sufficiently small and negative $\delta$.
For the loop black hole and complexification Type I and Type II respectively we find 
\be
&& \phi_{\mu}\phi^{\mu}  = - \frac{2 a_0^2 + a^2 r_*^2}{2 a^2 \delta ^2} - \frac{2 a r_*}{\delta} 
+ {\rm const.} + {\mathcal O} (\delta), \nonumber \\
&&  \phi_{\mu}\phi^{\mu}  = - \frac{a_0^2 + a^2 r_*^2}{2 a^2 \delta ^2} - \frac{2 a r_*}{\delta} + {\rm const.}
+ {\mathcal O} (\delta)
\ee
which are always negative for small values of $\delta$.
We conclude there are no CTC in the region around $r \approx 0$ and $\theta \approx \pi/2$
contrary to the classical Kerr space-time.

Also for the full-polymeric metric like for the semi-polymeric one we have a region, for negative 
values of $r$, where the norm $\phi^{\mu}\phi_{\mu}$ changes sign. We have a good improvement 
respect to the classical metric but we still have CTC curves.

\subsection{Horizon transitions} 
The metric we have introduced shown that there is, if we align the surface with $a$ as $z$ axis, a 
relation between the critical points of the surface $C(r, \theta; a)=0$, defined by 
(\ref{fullpolyhorizons}), and the transitions of the null surface. In the following we will refer to this surface as
\emph{critical surface}. 
In this section we will try to make 
this relation precise on a more general setting, relying only on the properties of the surface for a 
generic \emph{perturbation} of the critical surface. 
In particular we assume there is a situation similar to the one of the previous section: a rotating 
black hole not in vacuum according to the Einstein equations but that in the limit of some parameters 
going to zero (and that we consider small in general), the metric we obtain is Kerr. The previous 
example catches general peculiarities enough to generalize it. The only assumption is that such 
\emph{corrections} of the Kerr metric do not break the main symmetries, that is, the spacetime
has a symmetry axis along which rotations keep the metric invariant.

The first thing to note is that the surface is embedded in three dimensions and can be Taylor expanded around
the critical points:
\be 
&& C(r\approx r_c,\theta\approx \theta_c; a(r\approx r_c,\theta\approx \theta_c)) \nonumber \\
&& \hspace{-0.5cm} 
= C(r_c,\theta_c; a(r_c,\theta_c))+c_1 r^2 +c_2 \theta^2+ O(r^3, \theta^3), 
\ee
where $c_1$ and $c_2$, normalized to $\pm1$ are called the \emph{Morse indices} of the surface critical 
point. In general however we have to extend the notion of criticality also to points that cannot have an 
expansion of this form, like the points lying on $\theta=0$ or $\theta=\pi$, so to the notion of absolute minimum 
and absolute maximum.  Of course, if local critical points, there are three distinct possibilities: maxima, minima and saddle points corresponding 
to $(c_1=-1,c_2=-1)$,$(c_1=1,c_2=1)$ and $(c_1=1,c_2=-1)$ respectively. These indices are fundamental 
to study the topological properties of surfaces in Morse theory. However in our case the problem is 
slightly different due to the invariance of this surface respect with the $\phi$ variable, which 
corresponds to
 a global $\mathscr M\times\mathscr S$ topology, where $\mathscr M$ is the manifold given by the 
intersection of the $a={\rm const}.$ surface and the critical surface, as in Fig.\ref{fullhorizon}
and $\mathscr S$ is a circle.
In the following we will refer to \emph{tori} and \emph{spheres} to surfaces having the topology of tori 
and spheres respectively. It is easy to understand that in general local maxima coincide with 
\emph{shrinking}, saddle points with \emph{splitting}, and local minima \emph{create} spherical null surfaces, 
as the angular momentum, the parameter $a$, increases. Absolute minima
 and maxima occuring at $\theta_0=0$ and thus $\theta_0=\pi$ are slightly different as we explain now.
In fact the critical properties of the null surface is the global one; we have to translate the 
\emph{criticality} of $\mathscr M$ in the criticality of $\mathscr M \times \mathscr S$. The points $\theta=0$ 
and $\theta=\pi$ are different because they have to be identified, since they lie on the axis of symmetry; 
moreover due to the axial symmetry, critical points at an angle $0<\theta_0<\pi$ have an identical 
critical points at $\theta_0'=\pi-\theta_0$. 
By the Weierstress theorem to two local maxima coincide at least a minimum that, in this case, coincide with a 
saddle point
in between the two local maxima. In general local saddle points correspond to \emph{splits}: when the 
surface $a={\rm const}$ hit a saddle point
the null surface splits in two parts: at the angle where there is the saddle point the null surface bends and when the 
$a=const.$ surface hits the saddle point and the null surface separates. The way how this happens depends on the nature of the 
critical surface. We can apply the same analisys for local maxima. The surface 
at constant angular momentum cuts local maxima of the critical surface on cicles\footnote{With the topology of a circle.} if these are not at the extremal points, 
which correspond to a tori for the null surface, being $\mathscr S \times \mathscr S$. Thus local maxima 
in general correspond to 
\emph{shrinking} of tori or spheres if the maxima are at the extremal at increasing angular momentum. 
When the surface $a=const$ hits the critical point the tori shrinks to a point and the null surface 
disappears. The same happens for the maxima occurring at the extremal points. However a local minimum at 
the extremal points of the critical surface is tricky. In fact, when it occurs, the surfaces of the inner and outer horizons bend and touch 
on the axis. If, for example, the local minima at the extremal points are the first 
occurring the overall topology assumes a toroidal shape, transforming the inner and outer horizons in a torus. 
Local minima instead are the opposite of local maxima, when the surface $a=const$ hit them they create spherical null 
surfaces. This analysis complete the picture, in the approximation of little \emph{corrections} to the Kerr metric, 
of transitions from the inner-outer horizon to null-surfaces free spacetime. 

\section{Towards spinning LBH with charge}
In this section we consider a black with spin and electric charge. First we introduce 
the generalization of the Schwarzschild LBHs (semi-polymeric and full-polymeric) 
to the Reissner-Nordstr\"{o}m LBHs then we apply again the Newmann-Janis 
complex transformation to obtain rotating and  charged LBHs. 

\subsection{Reissner-Nordstr\"{o}m LBH}
It is easy to extend the spherically symmetric LBHs to the case of a black hole with charge.
We consider first the semi-polymeric case and then the full-polymeric case. 

\subsubsection{Semi-polymeric case} 
We recall the semi-polymeric metric for the spherically symmetric case without charge
\be
&& ds^2= 
\frac{r^2- 2 m r }{H(r)} dt^2 - \frac{dr^2}{\frac{r^2- 2 m r }{H(r)}}
- H(r) d \Omega^{(2)}, \nonumber \\
&& H(r) = r^2 + \frac{a_o^2}{r^2}.
\label{SP}
\ee
Is very simple to introduce the charge and to obtain a regular metric with the correct classical limit.
This can be done by the replacement \cite{RNamano}
\be
2 m r \rightarrow 2 m r - e^2
\ee
in (\ref{SP}), where $e$ is the electric charge. 
The metric is very close to the Reissner-Nordstr\"{o}m but with a bounce of the $S^2$ sphere
on a minimum area $a_o$ which solve the singularity problem. It is easy to show going through 
the analysis in \cite{Modesto:2008im} the regularity of the metric for any value, positive negative or zero,
of the radial coordinate.

\subsubsection{Full-polymeric case} 
The generalization of the full-polymeric LBH to a charged black hole 
is also very simple. We recall again the metric 
\be
&& \hspace{-0.3cm} ds^2 = g_{tt} dt^2 + g_{rr} dr^2 - H(r) d \Omega^{(2)}, \nonumber \\
&& \hspace{-0.3cm} g_{tt} 
= \frac{\left(r^2 - (r_+ + r_-) r   + r_+ r_-  \right)}{H(r)} \left( 1 + \frac{r_*}{r} \right)^2  , \nonumber \\
&& \hspace{-0.3cm}  g_{rr} := 
- \frac{   \, H(r) }{ \left( r^2 - (r_+ + r_-) r + r_+ r_-  \right)  }  \left( 1+\frac{r_*}{r} \right)^2,  \nonumber \\
&& \hspace{-0.3cm} 
H(r) = r^2 +\frac{a_o^2}{r^2} .
\label{componentsRN}
\ee
In this case we replace 
\be
&& (r_+ + r_-) r \rightarrow (r_+ + r_-) r - e^2, 
\label{RNhand}
\ee
The horizons are now located in 
\be
\hspace{-0.3cm} \tilde{r}_{\pm} = \frac{(r_+ + r_-) \pm \sqrt{ (r_+ + r_-)^2 - 4 (e^2 + r_+ r_-)}}{2}.
\ee
 It is easy to show going through 
the analysis in \cite{Modesto:2008im} the regularity of the metric for any value 
of the radial coordinate. Radial geodesic's analysis shows that we can not reach 
the center $r=0$ in finite time like in case of $e=0$.

For $m (1 - P)^2 \geqslant e$ we can express the metric in the following way
\be
&& \hspace{-0.3cm} g_{tt} 
= \frac{\left(r^2 - (\tilde{r}_+ + \tilde{r}_-) r   + \tilde{r}_+ \tilde{r}_-  \right)}{H(r)} \left( 1 + \frac{\tilde{r}_*}{r} \right)^2  , \nonumber \\
&& \hspace{-0.3cm}  g_{rr} := 
- \frac{   \, H(r) }{ \left( r^2 - (\tilde{r}_+ + \tilde{r}_-) r + \tilde{r}_+ \tilde{r}_-  \right)  }  \left( 1+\frac{\tilde{r}_*}{r} \right)^2,  \nonumber \\
&& \hspace{-0.3cm} 
H(r) =  r^2 +\frac{a_o^2}{r^2}  , 
\label{componentsRN2}
\ee
where we have introduced $\tilde{r}_* = \tilde{r}_+ \tilde{r}_-$ if we want to keep 
the duality property of the metric. 
It is easy to see that the Reissner-Nordstr\"{o}m full-polymeric LBH has exactly the same 
shape of the Schwarzschild full-polymeric LBH with $r_+$, $r_-$ and $r_*$ replaced by
$\tilde{r}_+$, $\tilde{r}_-$ and $\tilde{r}_*$ at least for $m (1 - P)^2 \geqslant e$; 
this makes very easy to derive the full-polymeric
Kerr-Newmann space-time.

\subsection{Kerr-Newmann LBH}
In this section we apply the Newmann-Janis complexification to the Reissner-Nordstr\"{o}m LBH
in its semi-polymeric and full-polymeric form. The following derivation is justified by the 
decoupling between polymerization of the space and electric field.
This is easy to see in the semi-polymeric case but a conjecture in the full-polymeric one.

\subsubsection{Semi-polymeric case} 
The complexification is straightforward and following section \ref{TypeI} the natural choose  
of $G$ and $H$ are 
\be
&& G(r) \rightarrow \frac{\rho^2 - 2 m r + e^2}{H(r)}, \nonumber \\
&& H(r) \rightarrow \rho^2 + \frac{a_o^2}{r^2}   \,\,\,\, {\rm for \,\, Type \,\, I} , \nonumber \\
&& H(r) \rightarrow \rho^2 + \frac{a_o^2}{\rho^2}   \,\,\,\, {\rm for \,\, Type \,\, II} ,
\ee
where $\rho^2 = r^2 + a^2 \cos^2 \theta$.
 The Kerr-Newmann LBH in Boyer-Lindquist coordinates reads 
\be
&&  \hspace{-0.7cm}  ds^2 =  \frac{\Delta - a^2 \sin^2 \theta}{\Sigma} dt^2 - \frac{ \Sigma}{\Delta }  \, dr^2    \nonumber \\
&& \hspace{-0.7cm} 
 +  2 a \sin^2 \theta \left(1 - \frac{\Delta - a^2 \sin^2 \theta}{\Sigma} \right) dt \, d \phi  
 - \Sigma \, d \theta^2 
 \nonumber \\
&&  \hspace{-0.7cm} 
 -  \, \sin ^2 \theta  \left[ \Sigma + a^2 \sin^2 \theta \left(2 - \frac{\Delta - a^2 \sin^2\theta}{\Sigma}\right)   \right]    d \phi^2
%
%
  \label{BHKNLBH}
\ee
where now $\Delta(r)$ is 
\be
\Delta(r) = r^2 - 2 m r + e^2 + a^2.
\label{deltaKN}
\ee
The rest of the analysis follows exactly section \ref{TypeI} and \ref{TypeII} but with the new $\Delta(r)$
function defined in (\ref{deltaKN}). 
Particularly for the Type I complexification, $\Omega_H$, $\kappa_{\pm}$ and 
the event horizon area have the same shape as those calculated in section \ref{TypeI} but with $r_{\pm}$ replaced 
with the roots of the new equation $\Delta(r) = 0$ defined in (\ref{deltaKN}).

\subsubsection{Full-polymeric case} 
This section is very short since it is identical to \ref{FullPolyKerr}
if we replace everywhere  $r_+$, $r_-$ and $r_*$ with 
$\tilde{r}_+$, $\tilde{r}_-$ and $\tilde{r}_*$ when $m (1 - P)^2 \geqslant e$. 
If $m (1 - P)^2 < e$ we still define $\tilde{r}_*^2 = \tilde{r}_+ \tilde{r}_- = e^2 + r_+  r_-$ but 
we apply the Newmann-Janis transformation directly to (\ref{componentsRN}) 
with the replacement  (\ref{RNhand}).

\section{Conclusions}

In this paper we used the Newman-Janis algorithm to construct 
regular spinning black hole from the Schwarzschild loop black hole. 
We used constraints coming from the classical limits and arguments
from the Newman-Janis algorithm applied in the past to the Schwarzshild metric and The Reissner-Nordstr\"{o}m metric. 
We found {\em Kerr-like geometries without ring singularity}. 
These results, while not definitive,
hints in the direction that the polimeric quantization inspired by loop quantum gravity could solve the singularity problem also for the Kerr spacetime. 
We started considering two different spherically symmetric space-time obtained in 
\cite{Modesto:2008im} that we called semi-polymeric and full-polymeric.
The first metric can be obtained from the second one in an appropriate limit.
We studied the semi-polymeric one for reasons of pure simplicity since such metric 
has all good property of regularity. In the paper we introduced also 
the notation Type I and Type II to indicate the two complexifications we used.
For the semi-polymeric spinning loop black hole we showed explicitly that the Ricci scalar 
and the Kretschmann invariant are regular in $r = 0$ and $\theta = \pi/2$ (for the semi-polymeric 
Type I loop black hole the reader can fine the explicit formula for the Ricci scalar and the Kretschmann
invariant). The structure of the event horizon and of the ergosphere is the same of the classical
Kerr metric and the causal space-time structure is given in the text for each case in terms of 
Penrose diagrams.
The full-polymeric spinning loop black hole has a more reach structure. The ergosphere surfaces 
are very similar to the classical ones but the horizon surfaces, while are very similar to the classical ones
for small values of the angular momentum, change topology for large value of the angular momentum 
respect to the mass. 
The singularity here is also cured but in a more elegant way. Any observer in the equatorial plane
($\theta = \pi/2$) never can reach the point $r =0$ starting from positive or negative values of the radial coordinate. Of course the Ricci scalar and the Kretschmann 
invariant are regular.

For the first time we introduce the Reissner-Nordstr\"{o}m LBH metric and extended the Newmann-Janis 
transformation to this to obtain the Kerr-Newmann LBH with spin and electric charge.
The properties of the spinning LBHs are shared by the spinning and charged LBHs.
For all semi-polymeric cases studied there are no naked singularities for any value of the angular momentum.

We studied the presence of CTCs (closed timelike curves) in the region 
near $r \approx 0$ and $\theta \approx \pi/2$ and we have shown CTCs disappear in 
all the new metrics. In particular for the full polymeric since each observer never can arrive 
in $r=0$ there is no physical reason to extend the space-time to negative values of $r$ where
classically  the CTCs are located. This result does not exclude the existence of other CTC's regions 
for negative values of $r$ where $g_{\phi \phi}$ changes sign. 

In this paper we did not solve the equations of motion coming from a fundamental theory 
but we simply introduced the angular momentum in spherically symmetric solutions
by the Newmann-Janis transformation.
However we can always see spherically symmetric LBHs to be solutions of the Einstein 
theory with an effective energy tensor: $G_{\mu \nu} = 8 \pi T^{\rm QG}_{\mu\nu}$;
where $T^{\rm QG}_{\mu \nu}$ summarizes the {\em loop corrections}. 
The {Spinning LBHs} obtained in this paper are actually solutions of the Einstein equations 
with a stress energy tensor obtained from the spherically symmetric one applying 
the Newmann-Janis transformation properly. 
The effective stress energy tensor is a function of two or three parameters depending 
on the semi-polymeric or full-polymeric nature of the LBH:
 \begin{eqnarray}
G_{\mu \nu} = 8 \pi \,    \left\{ \begin{array}{ll} 
       T^{\rm QG}_{\mu \nu}(a , P) \, , \,\,\,\,\,\,\,\,\,\,\,  {\rm SEMI-POLYMERIC} \, ,
        \vspace{0.04cm}    \\ 
             T^{\rm QG}_{\mu \nu}(a , P, a_o) \, , \,\,\, {\rm FULL-POLYMERIC}.
        \end{array} \right. \nonumber 
\end{eqnarray}
We conclude the paper summarizing the metrics obtained. 
%
\begin{widetext}
\begin{center}
SEMI-POLYMERIC :
\end{center}
\be
&& ds^2 =  \frac{\Delta - a^2 \sin^2 \theta}{\Sigma} dt^2 - \frac{ \Sigma}{\Delta }  \, dr^2  
 - \Sigma \, d \theta^2 
 \nonumber \\
&&  \hspace{-0.7cm} +  2 a \sin^2 \theta \left(1 - \frac{\Delta - a^2 \sin^2 \theta}{\Sigma} \right) dt \, d \phi  
 -  \, \sin ^2 \theta  \left[ \Sigma + a^2 \sin^2 \theta \left(2 - \frac{\Delta - a^2 \sin^2\theta}{\Sigma}\right)   \right]    d \phi^2, \nonumber \\
&& \Delta(r) = r^2 - 2 m r +a^2\, . \nonumber 
 \ee
\begin{center}
FULL-POLYMERIC :
\end{center}
\be
&&ds^2 =   \frac{ \left( \Delta - a^2 \sin^2 \theta \right) (\rho^2 + r r_* )^2}{\rho^4 \, \Sigma} dt^2  
 - \frac{ \Sigma \, (\rho^2 + r r_*)^2}{\rho^4  \left( \Delta - a^2 \sin^2 \theta \right)
+ a^2 \sin^2 \theta ( \rho^2 + r r_*)^2} dr^2 - \Sigma d \theta^2  \nonumber \\
&& \hspace{-0.4cm} 
+2 \frac{a \sin^2 \theta \left( \rho^2 + r r_* \right)^2 \left[ \Sigma - 
( \Delta - a^2 \sin^2 \theta  ) \right]}{\Sigma \, \rho^4} dt d \phi  
 -  \sin^2 \theta \Big[ \Sigma 
 + a^2 \sin^2 \theta \frac{ ( \rho^2 + r r_* )^2 
\left( 2 \Sigma - (\Delta - a^2 \sin^2 \theta  ) \right) }{\Sigma \, \rho^4} \Big] d \phi^2,
\nonumber \\
&& \Delta = r^2 - (r_+ + r_-) r + r_+ r_-  +a^2 \, , \nonumber \\
&& {\rm Type \,\, I} \, : \,\,\,\,  \Sigma = r^2 + a^2 \cos^2 \theta +\frac{a_o^2}{r^2}, \nonumber \\
&& {\rm Type \,\, II} \, : \,\,\,\,  \Sigma = r^2 + a^2 \cos^2 \theta +\frac{a_o^2}{r^2 + a^2 \cos^2 \theta}. \nonumber 
\ee
\end{widetext}
The function $\Sigma$ is the same for both the metrics.

\begin{acknowledgments}
\noindent The authors are indebted with S.P. Drake for clarifications on the Newman-Janis algorithm
and Roberto Balbinot for the useful comments on the black hole thermodynamics.
Research at Perimeter Institute is
supported by the Government of Canada through Industry Canada and by the Province of Ontario through the
Ministry of Research \& Innovation.
\end{acknowledgments}

\section{Appendices}

\subsection{Kretschmann invariant for the Type I semi-polymeric LBH}
The Kretschmann invariant for the Type I complexified metric reads
\begin{eqnarray}
&& \hspace{-0.2cm} 
K(r, \theta) = 
\frac{16}{\left(2 r^4+a^2 r^2+a^2 \cos (2 \theta ) r^2+2 a_o^2\right)^6} \times \nonumber \\
&& \hspace{-0.2cm} 
(192 m^2 r^{18}-1440 a^2 m^2 r^{16}+1536 a_o^2 m r^{15}
+1080 a^4 \nonumber 
\ee
\be
&&  \hspace{-0.2cm}
m^2 r^{14}-3072 a_o^2 m^2
   r^{14}
   -2880 a^2 a_o^2 m r^{13} +4032 a_o^4 r^{12} \nonumber \\
   && \hspace{-0.2cm}
   -60 a^6 m^2 r^{12}+7296 a^2 a_o^2 m^2 r^{12}  - 6 a^6 m^2
   \cos (6 \theta ) r^{12}      \nonumber \\
   && \hspace{-0.2cm}
  -17408 a_o^4 m r^{11}  -1536 a^4 a_o^2 m r^{11}   +5328 a^2 a_o^4 r^{10} \nonumber \\
&&   \hspace{-0.2cm}
+18816
   a_o^4 m^2 r^{10}-288 a^4 a_o^2 m^2 r^{10}  -10464 a^2 a_o^4 m r^9 \nonumber  \\
   &&  \hspace{-0.2cm}
    -312 a^6 a_o^2 m r^9+12 a^6 a_o^2 m \cos (6 \theta ) r^9 -2176 a_o^6 r^8  \nonumber \\
    && \hspace{-0.2cm}
   +3192 a^4 a_o^4 r^8-160 a^2 a_o^4 m^2 r^8 
   +7168
   a_o^6 m r^7   -3360   \nonumber \\
  && \hspace{-0.2cm}
   a^4 a_o^4 m r^7-672 a^2 a_o^6 r^6+1002 a^6 a_o^4 r^6-4608 a_o^6 m^2 r^6 \nonumber \\
   &&  \hspace{-0.2cm}
   +336 a^4 a_o^4 m^2 r^6 
   +3 a^6 a_o^4 \cos (6 \theta ) r^6-448 a^2 a_o^6 m r^5 \nonumber \\
   &&  \hspace{-0.2cm}
     -396 a^6 a_o^4 m r^5-18 a^6
   a_o^4 m \cos (6 \theta ) r^5+704 a_o^8 r^4 + 144 \nonumber \\
   &&   \hspace{-0.2cm}
    a^4 a_o^6 r^4 +144 a^8 a_o^4 r^4  
   +64 a^2 a_o^6
   m^2 r^4+9 a^8 a_o^4 \cos (6 \theta ) r^4 \nonumber \\
   &&  \hspace{-0.2cm}
   -1536 a_o^8 m r^3+88 a^4 a_o^6 m r^3-176 a^2 a_o^8 r^2+48
   a^6 a_o^6 r^2  \nonumber \\
   && \hspace{-0.2cm}
   +960 a_o^8 m^2 r^2-32 a^2 a_o^8 m r   
     +34 a^4 a_o^8+a^2 (207 a_o^4 r^4 a^6 \nonumber \\
     && \hspace{-0.2cm}
     +3
   (-30 m^2 r^{12}-132 a_o^2 m r^9+3 a_o^4 (133 r-62 m) r^5  \nonumber \\
   &&  \hspace{-0.2cm}
   +16 a_o^6 r^2) a^4 
   + 8 (180 m^2
   r^{14}-48 a_o^2 m (m+4 r) r^{10}   +2 a_o^4  \nonumber \\
   &&  \hspace{-0.2cm}
    (  28 m^2-216 r m+123 r^2) r^6+4 a_o^6 (m-3 r) r^3    +5    a_o^8  ) a^2    \nonumber \\
   &&    \hspace{-0.2cm}
    -16 r (90 m^2 r^{15}+12 a_o^2 m (21 r-38 m) r^{11}  +a_o^4 (10 m^2   \nonumber \\
    && \hspace{-0.2cm}
  +26 r m-     27   r^2) r^7-2 a_o^6 \left(2 m^2+2 r m+r^2\right) r^3  +a_o^8  \nonumber \\
&&   \hspace{-0.2cm}
(30 m -19 r) ) ) \cos (2 \theta )+2 a^4
  (18 m^2 \left(10 r^2-a^2\right) r^{12} \nonumber \\
  && \hspace{-0.2cm}
  -12 a_o^2 m \left(3 a^2+4 m r\right) r^9+a_o^4 (36
   a^4+\left(99 r^2-90 m r \right) a^2 \nonumber \\
&&   \hspace{-0.2cm}
+4 r^2 (  14 m^2-12 r m+9 r^2   ) ) r^4-4 a_o^6 (7 m-6 r) r^3 \nonumber \\
&& \hspace{-0.2cm}
+11
   a_o^8 ) \cos (4 \theta )). 
   \label{Kre}
\end{eqnarray}
This quantity is regular and finite everywhere and in particular 
\be \hspace{-0.2cm} 
\lim_{r \rightarrow 0} \Big( \lim_{ \theta \rightarrow \pi/2}  K(r, \theta)   \Big) = 
\lim_{\theta \rightarrow \pi/2 } \Big( \lim_{r \rightarrow 0}   K(r, \theta)  \Big) = \frac{4 a^4}{a_o^4}.
\ee
%
%

\subsection{Tortoise coordinates for the full polymeric metric}
The case of two distinct horizons with Type I complexification :
\be  
&& \hspace{-0.6cm} r^* = r -\frac{a_o^2}{r r_2 r_1} -
\frac{\left(a^2 r_2^2+a_o^2+r_2^4\right) }{r_2^2
   (r_1-r_2)}   \log | r-r_2 |   \\
   &&\hspace{-0.6cm}
   +\frac{\left(a^2 r_1^2 + a_o^2 + r_1^4\right)}{r_1^2
   (r_1-r_2)}   \log | r-r_1 | 
   +\frac{ \left(a_o^2 r_2+a_o^2 r_1\right)}{r_2^2
   r_1^2} \log | r |, \nonumber
\ee
where $r_{1,2}$ are the bigger and the smaller horizons for $\theta = 0$ and then 
coincide with the roots in (\ref{ergo1}).

The case of no horizons at all and Type I complexification:
\be 
&& r^* = r
-\frac{a_o^2}{r( a^2  +  r_- r_+)}+\frac{a_o^2 \log (r) (r_-+r_+)}{\left(a^2+r_-
   r_+\right)^2} \nonumber \\
   &&  \hspace{1cm}
   +\frac{(r_- + r_+) \left(a^4+2 a^2 r_- r_+-a_o^2+r_-^2
   r_+^2\right) }{2 \left(a^2+r_-
   r_+\right)^2} \times \nonumber \\
   && \hspace{1cm}
   \log \left(a^2+(r-r_-) (r-r_+)\right) \nonumber \\
   %
  && \hspace{1cm} + \frac{\arctan \left[\frac{2 r-r_--r_+}{\sqrt{4 a^2-(r_- - r_+)^2}}\right]}{\sqrt{4 a^2-(r_- - r_+)^2} \left(a^2+r_- r_+\right)^2} \times \nonumber \\
  && \hspace{1cm}
   \Big[a^4 \left(r_-^2+r_+^2\right)+a^2 \left(2 r_- r_+
   \left(r_-^2+r_+^2\right)-2 a_o^2\right)
   \nonumber \\
   && \hspace{1cm} +\left(r_-^2+r_+^2\right)
   \left(a_o^2+r_-^2 r_+^2\right)  \Big]. 
\ee


\end{document}